\documentclass[preprint,12pt,3p]{elsarticle}
\usepackage{amssymb}
\usepackage{subfigure}
\usepackage{amsmath}
\journal{}

\begin{document}

\begin{frontmatter}

\title{Generalized Collision-free Velocity Model for\\Pedestrian Dynamics}

\author[label1]{Qiancheng Xu\corref{cor1}}
\address[label1]{Institute for Advanced Simulation,\\ Forschungszentrum J\"ulich GmbH, 52425 J\"ulich, Germany}

\cortext[cor1]{I am corresponding author}

\ead{q.xu@fz-juelich.de}

\author[label1]{Mohcine Chraibi}
\ead{m.chraibi@fz-juelich.de}

\author[label2]{Antoine Tordeux}
\address[label2]{School of Mechanical Engineering and Safety Engineering,\\University of Wuppertal, 42119 Wuppertal, Germany}
\ead{tordeux@uni-wuppertal.de}

\author[label3]{Jun Zhang}
\address[label3]{State Key Laboratory of Fire Science, \\University of Science and Technology of China, 230027 Hefei, China}
\ead{junz@ustc.edu.cn}

\begin{abstract}
The collision-free velocity model is a microscopic pedestrian model, which despite its simplicity, reproduces fairly well several self-organization phenomena in pedestrian dynamics. 
The model consists of two components: a direction sub-model that combines individual desired moving direction and neighbour's influence to imitate the process of navigating in a two-dimensional space, and an intrinsically collision-free speed sub-model which controls the speed of the agents with respect to the distance to their neighbors.

This paper generalizes the collision-free velocity model by introducing the influence of walls and extending the distance calculations to velocity-based ellipses.
Besides, we introduce enhancements to the direction sub-module of the model that smooth the direction changes of pedestrians in the simulation; a shortcoming that was not visible in the original model due to the symmetry of the circular shapes. 
Moreover, the introduced improvements mitigate backward movement, leading to a more realistic distribution of pedestrians especially in bottleneck scenarios.

We study by simulation the effects of the pedestrian's shape by comparing the fundamental diagram in narrow and wide corridors. 
Furthermore, we validate our enhancements to the direction sub-model for bottleneck experiments with varying  exit's width. 
\end{abstract}


\end{frontmatter}


\section{Introduction}
Nowadays, the scale of crowd activities is getting bigger with the constant increase of the world population and the convenience of transport. 
Although these events usually are carefully planned before they are held, the probability of accidents cannot be neglected, especially when the number of participants is considerably high.
Besides, in some complex buildings, such as train stations, airports, stadiums, commercial malls, crowd density can be relatively high, in particular during rush hours. 
For increasing the comfort and usability of these facilities, simulations of pedestrian dynamics may help during the design of buildings and even after their construction to identify potential bottlenecks and mitigate their effects \cite{review1,review2,review3}.
 
In general, models used to describe pedestrian dynamics can be categorized as macroscopic models, mesoscopic models and microscopic models.
Macroscopic models  \cite{macro1,macro2,macro3,macro4} rely on aggregated quantities  e.g.\ density, velocity and flow to describe pedestrian dynamics. 
The intermediate scale between the other two classes is mesoscopic models \cite{meso0,meso1,meso2,meso3} which borrows the idea from kinetic theory to describe the crowds as systems of probability density functions for `active particles'. 
Generally speaking, microscopic models consider each pedestrian's movement individually, mesoscopic models describe their motion in stochastic continuum while macroscopic models deal with the averaged performances. 

The model proposed in this paper belongs to the microscopic model category. 
We will introduce this class in detail in the following.

Compared to macroscopic and mesoscopic models, microscopic models are often more complex and may be computationally expensive, but can represent the behaviour of pedestrians in more details.
After more than 50 years of development, many kinds of microscopic models exist in the literature. 
Most of the models can reproduce fairly well several collective phenomena in pedestrian dynamics \cite{review2}. 
We can distinguish between cellular automate models \cite{ca1,ca2,ca3,ca4,ca5} (0th order models), velocity models \cite{vm1,vm2Cvm,vm3,vm4} (1st order models) and force-based models \cite{sfm1,mohcine,sfm3,sfm4} (2nd order models). 
While the former models are discrete in space and computationally fast, the later models are continuous in space and hence are easier to use in complex geometries. 
Whether continuous models are computationally expensive depends not only on the order of the model, but also on it’s definition. 
However, generally speaking, first-order models are less expensive since their numerical solution involves only one integration step, while two integration steps are required for second-order models. 
Furthermore, highest discretisation levels generally require stronger smallness conditions for the simulation time step to get numerical stability.
In this paper, we focus on the extension of the collision-free velocity model introduced in \cite{vm2Cvm}.
 
The collision-free velocity model (CVM)~\cite{vm2Cvm} is a velocity-model, composed of a speed and a direction sub-models. 
Unlike most force-based models, CVM, being a first-order model, is per definition collision-free. 

In this paper, we extend the CVM by considering the influence of walls and integrating two extensions for the minimal model.
First, we change the shape of agents from circle to dynamical ellipse. 
In the original model, circles are used to express the projection of the pedestrian's body on the two-dimensional plane.
However, many references and researches indicate that dynamical ellipse can represent pedestrian's shape more accurately since the space a pedestrian occupies is influenced by the length of the legs during the motion and the lateral swaying of the body \cite{mohcine}. 
Therefore, we generalize CVM by extending the distance calculation to velocity-based ellipse and compare the simulation results with the original model (circles).
After introducing the first extension, an unnatural ``shaking'' was observed during the simulation, which is caused by the zero-order direction sub-model for the direction. 
We propose a new first-order direction sub-model, designed to smooth the direction changes of pedestrians in the simulation.
 
For the sake of completeness, we briefly introduce the original CVM in section~\ref{sec:sfvm}.
The generalization of the model from circle-based to a ellipse-based definition and the new direction sub-model are given in  section~\ref{sec:extension}. 
In section~\ref{sec:simulation}, the comparison between the simulation results of a circle and a velocity-based ellipse is given and the performances of new direction sub-model are compared to the original CVM.
Finally, we give a summary of the extensions and discuss limitations of the model as well as future research directions in the concluding section~\ref{sec:conclusion}.

\section{Collision-free velocity model}
\label{sec:sfvm}
\begin{figure}[htbp]
    \centering
    \includegraphics[width=0.7\textwidth]{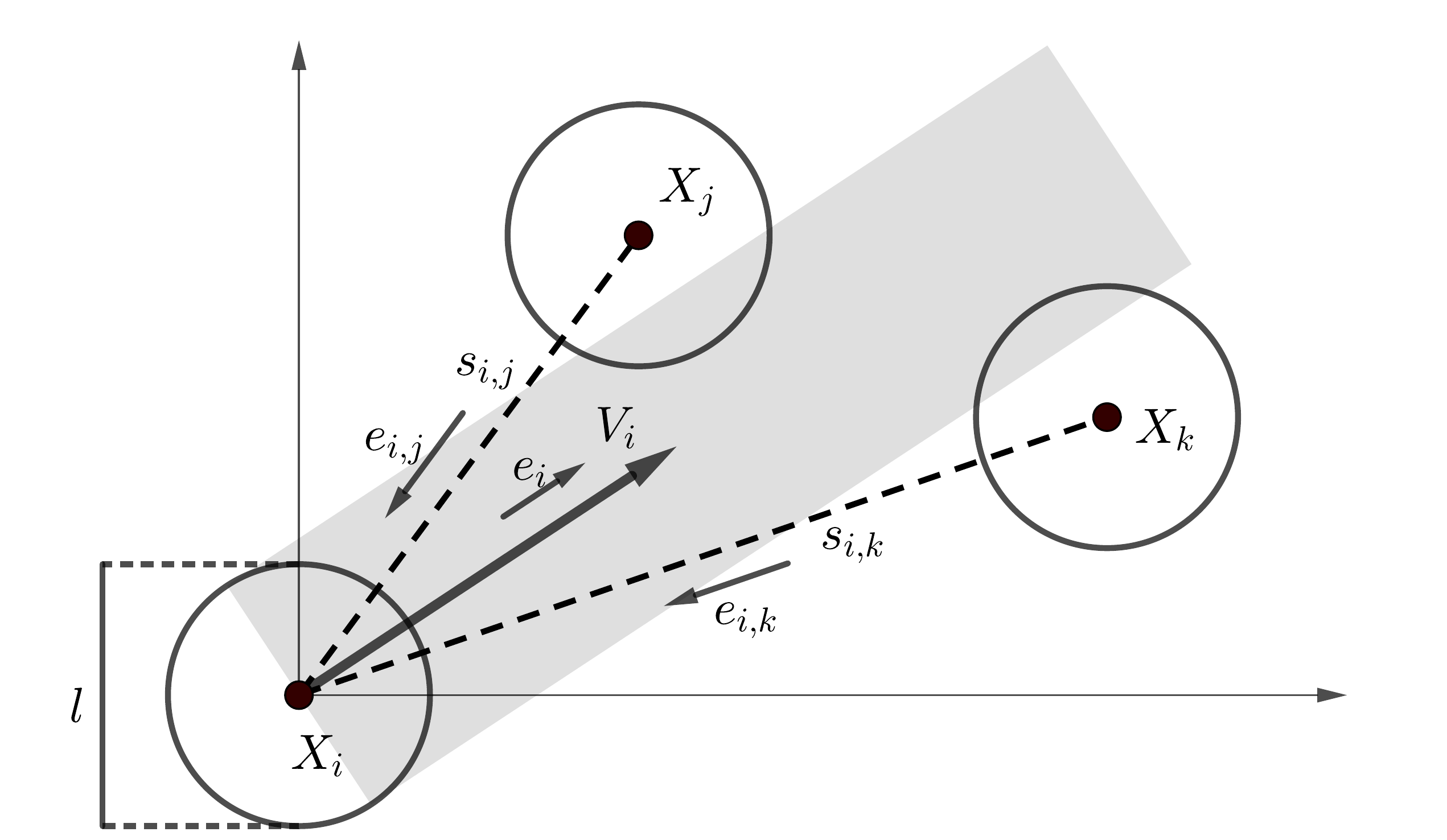}
    \caption{Notations used in the collision-free velocity model. 
    $\ell$ is the size of agents, $X_i$, $X_j$ and $X_k$ are positions of pedestrians, $\vec{e}_i$ is the moving direction of pedestrian $i$ and $V_i$ is the  moving speed, $s_{i,j}$ and $s_{i,k}$ are distances between the centers of pedestrians, $\vec{e}_{i,j}$ and $\vec{e}_{i,k}$ and the unit vector from $X_j$ and $X_k$ to $X_i$.   }
    \label{fig:Original model}
\end{figure}

In the original model, the moving direction and velocity of each pedestrian are updated at each time step. 
Moving direction of a pedestrian is obtained by superposing the influence of the surrounding pedestrians and the desired moving direction. 
The value of the speed depends on the minimum spacing in the moving direction. 
In figure \ref{fig:Original model} (borrowed from \cite{vm2Cvm}), pedestrians are modeled as circles with constant diameter $\ell$.
$X_i$, $X_j$ and $X_k$ are positions of pedestrians $i$, $j$ and $k$. 
The original CVM is described as
\begin{equation}
    \dot{X}_i(X_i,X_j,\dots)=V_i(X_i,X_j,\dots)\cdot\vec{e}_i(X_i,X_j,\dots),
\end{equation}
where $V_i$ is the speed of pedestrian $i$ and $\vec{e}_i$ is the moving direction.

Moving direction $\vec{e}_i$ is obtained from the direction sub-model
\begin{equation}
   \vec{e}_i(X_i,X_j,...)=u_1\cdot\Big(\vec{e}_i^{\ 0}+\sum_{j\in N_i} R(s_{i,j})\cdot \vec{e}_{i,j}\Big),
\end{equation}
where $u_1$ is a normalization constant such that $||\vec{e}_i||=1$, $\vec{e}_i^{\ 0}$ is the desired direction towards a certain goal, $N_i$ is the set containing all the neighbours of the pedestrian $i$, $\vec{e}_{i,j}$ is the unit vector from the center of the pedestrian $j$ towards the center of the pedestrian $i$. 
The function
\begin{equation}
    R(s_{i,j})=k\cdot \exp\Big(\frac{\ell-s_{i,j}}{D}\Big),
\end{equation}
is used to describe the influence that neighbours act on the moving direction of pedestrian $i$.  
The strength coefficient $k>0$ and the distance coefficient $D>0$ calibrate the function accordingly. 
As mentioned before, $\ell$ is the diameter of the circle used to represent the pedestrians and $s_{i,j}$ is the distance between the centers of pedestrian $i$ and $j$.

After obtaining the moving direction $\vec{e}_i$, the speed model 
\begin{equation}
\label{equ:original velocity function}
    V_i(s_{i,j})=\min\Big\{V_{i}^0,\max\big\{0,\frac{s_{i}-\ell}{T}\big\}\Big\},
\end{equation}
is used to determine the scale of velocity $V_i$ in the direction $\vec{e}_i$. 
In Eq.~(\ref{equ:original velocity function}), $V_{i}^0$ is the desired speed of pedestrian $i$ and
\begin{equation}
\label{equ:spacing in front i}
    s_i=\min_{j\in J_i}s_{i,j},
\end{equation}
is the distance between the center of pedestrian $i$ and the center of the closet pedestrian in front of pedestrian $i$, when pedestrian $i$ moving in the direction $\vec{e}_i$.
The definition of set $J_i$ in Eq.~(\ref{equ:spacing in front i}) is
\begin{equation}
    J_i=\Big\{j, \vec{e}_i \cdot \vec{e}_{i,j} \leq 0\ \text{and}\ \left|\vec{e}_i^\bot \cdot \vec{e}_{i,j}\right| \leq \frac{\ell}{s_{i,j}}\Big\},
\end{equation}
where $\vec{e}_i^\bot \cdot \vec{e}_i =0 $.
$J_i$ is the set of all pedestrians overlapping with the grey area in figure \ref{fig:Original model}. 
The only coefficient in the speed model is $T>0$ which is used to adjust the gap between pedestrians. 

\begin{figure}[htbp]
    \centering
    \includegraphics[width=0.7\textwidth]{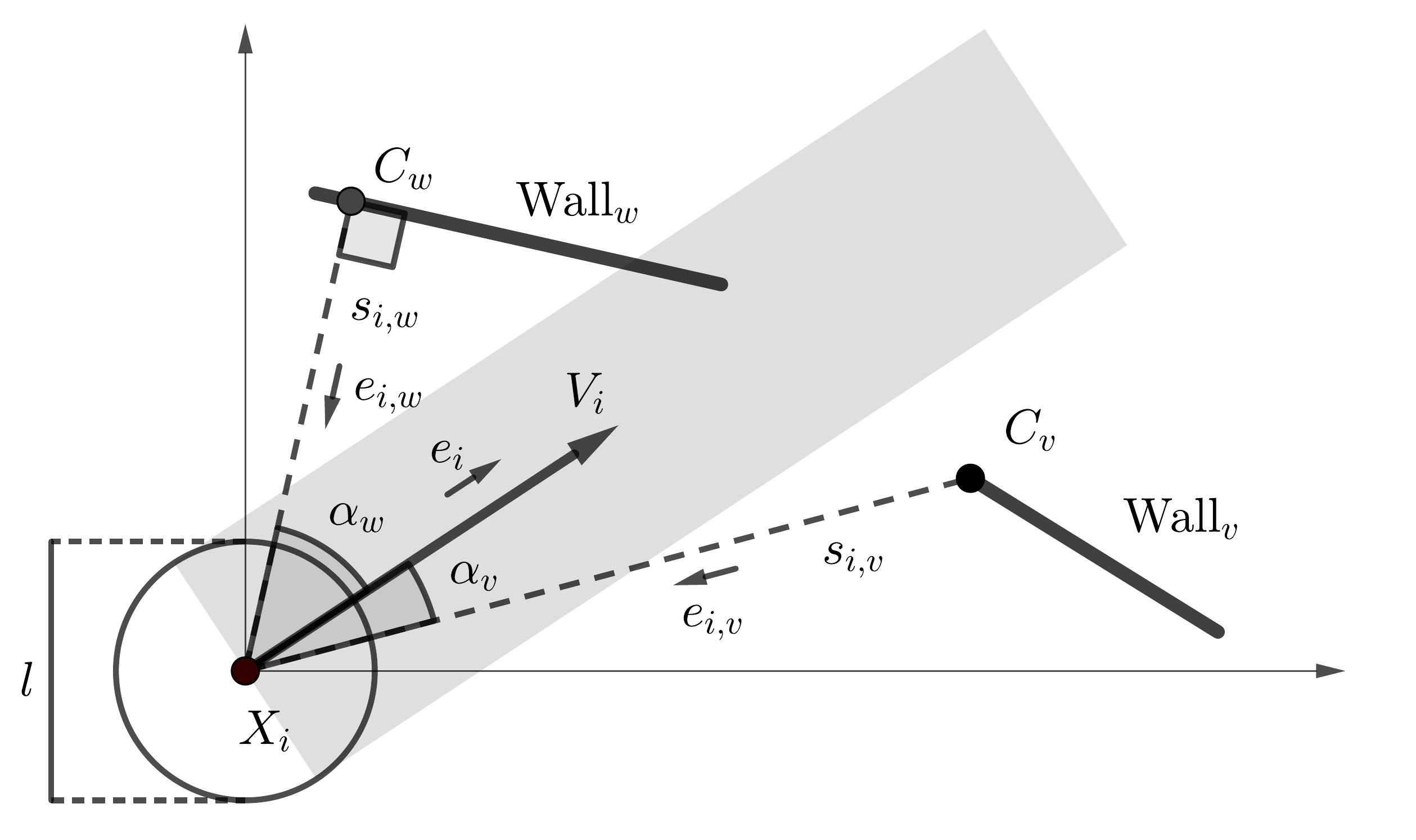}
    \caption{Notations used in the collision-free velocity model when calculating the influence of walls. 
    $\ell$, $X_i$, $\vec{e}_i$, $V_i$ are the size, position, moving direction and moving speed of pedestrian $i$, $C_v$ and $C_w$ are the closest points in wall $v$ and $w$ to $X_i$, $\vec{e}_{i,v}$ and $\vec{e}_{i,w}$ are the unit vectors from $C_v$ and $C_w$ to $X_i$, $s_{i,v}$ and $s_{i,w}$ are the distances from $C_v$ and $C_w$ to $X_i$, $\alpha_v$ is the angle between $\vec{e}_i$ and $-\vec{e}_{i,v}$, $\alpha_w$ is the angle between $\vec{e}_i$ and $-\vec{e}_{i,w}$.}
    \label{fig:wallinfluence}
\end{figure}

The above-mentioned definition of the CVM, describes specifically interactions among pedestrians. 
However, the influence of walls and obstacles has been left from the definition of the model.
In this work, we close this gap by only considering straight walls. 
If the shape of the wall in the simulation is irregular, then we will approximate it to a few straight walls.
In figure~\ref{fig:wallinfluence}, $X_i$, $\vec{e}_i$ and $V_i$ have the same definitions as in figure~\ref{fig:Original model}. 
Besides, there are two walls in the figure, wall $v$ and $w$. 
$C_v$ and $C_w$ are the closest points in wall $v$ and $w$ to the center of pedestrian $i$ respectively. 
$\vec{e}_{i,v}$ and $\vec{e}_{i,w}$ are the unit vectors from $C_v$ and $C_w$ to $X_i$. 
$s_{i,v}$ and $s_{i,w}$ are the distances from $C_v$ and $C_w$ to $X_i$. 
The angle between $\vec{e}_i$ and $-\vec{e}_{i,v}$ is $\alpha_v$ and the angle between $\vec{e}_i$ and $-\vec{e}_{i,w}$ is $\alpha_w$. 

After introducing the influence of walls, the direction model becomes
\begin{equation}
\label{equ:direction model with wall}
   \vec{e}_i=u_2\cdot\Big(\vec{e}_i^{\ 0}+\sum_{j\in N_i} R(s_{i,j})\cdot \vec{e}_{i,j}+\sum_{v\in W_i} R_w(s_{i,v})\cdot \vec{e}_{i,v}\Big),
\end{equation}
where $u_2$ is a normalization constant such that $||\vec{e}_i||=1$, $W_i$ is the set of walls nearby pedestrian $i$, and 
\begin{equation}
    R_w(s_{i,v})=k_w\cdot \exp\Big(\frac{\frac{\ell}{2}-s_{i,v}}{D_w}\Big),
\end{equation}
where $k_w>0$ and $D_w>0$ are used to calibrate the function accordingly.

In order to avoid overlaps of pedestrians with walls, walls should not only influence pedestrian's moving direction but also their speed. 
The expanded speed model is
\begin{equation}
    V_i=\min\Big\{V_{i}^0,\max\big\{0,\frac{s_{i}-\ell}{T}\big\},\max\big\{0,\frac{sw_{i}}{T}\big\}\Big\},
\end{equation}
where the definitions of $s_i$, $\ell$, $T$ are same as in Eq.~(\ref{equ:original velocity function}) and 
\begin{equation}
    sw_i=\min_{v\in JW_i} \frac{s_{i,v}-\frac{\ell}{2}}{\cos{\alpha_v}},
\end{equation}
where $JW_i$ is the set containing all the walls in the moving direction of pedestrian $i$ (grey area in figure \ref{fig:wallinfluence}). 

\section{Generalization of the collision-free velocity model}
\label{sec:extension}
In this section we introduce extensions of the CVM.
We also show how every extension influences the resulting dynamics and eventually enhances the simulation results.

\subsection{From circle to ellipse}
We generalize the collision-free velocity model by extending the distance calculations to velocity-based ellipses.
The plane view of the pedestrian $i$'s body is represented by an ellipse \cite{ellipse}.
The major semi-axis $a_i$ and minor semi-axis $b_i$ of the ellipse represent the space requirement in the direction of motion and along the shoulder axis respectively. 

In~\cite{mohcine} the semi-axis along the walking direction is defined as 
\begin{equation}
\label{equ:a}
    a_i=a_{\min}+\tau_a V_i,
\end{equation}
where $V_i$ is the speed of pedestrian $i$, while $a_{\min}>0$ and $\tau_a>0$ are two parameters.

The idea that the semi-axis of the ellipse along the walking direction vary with speed is derived from the fact that the spacing a pedestrian needed in his/her moving direction has a positive correlation with his/her speed \cite{aminT}.
This in turn is also the role of parameter $T$ which is defined in the speed sub-model to adjust the gap between agents.
We conclude that in our model $T$ and $\tau_a$ model the same behavior of pedestrians even if their physical interpretations are different.
 
\begin{figure}[htbp]
    \centering
    \includegraphics[width=0.7\textwidth]{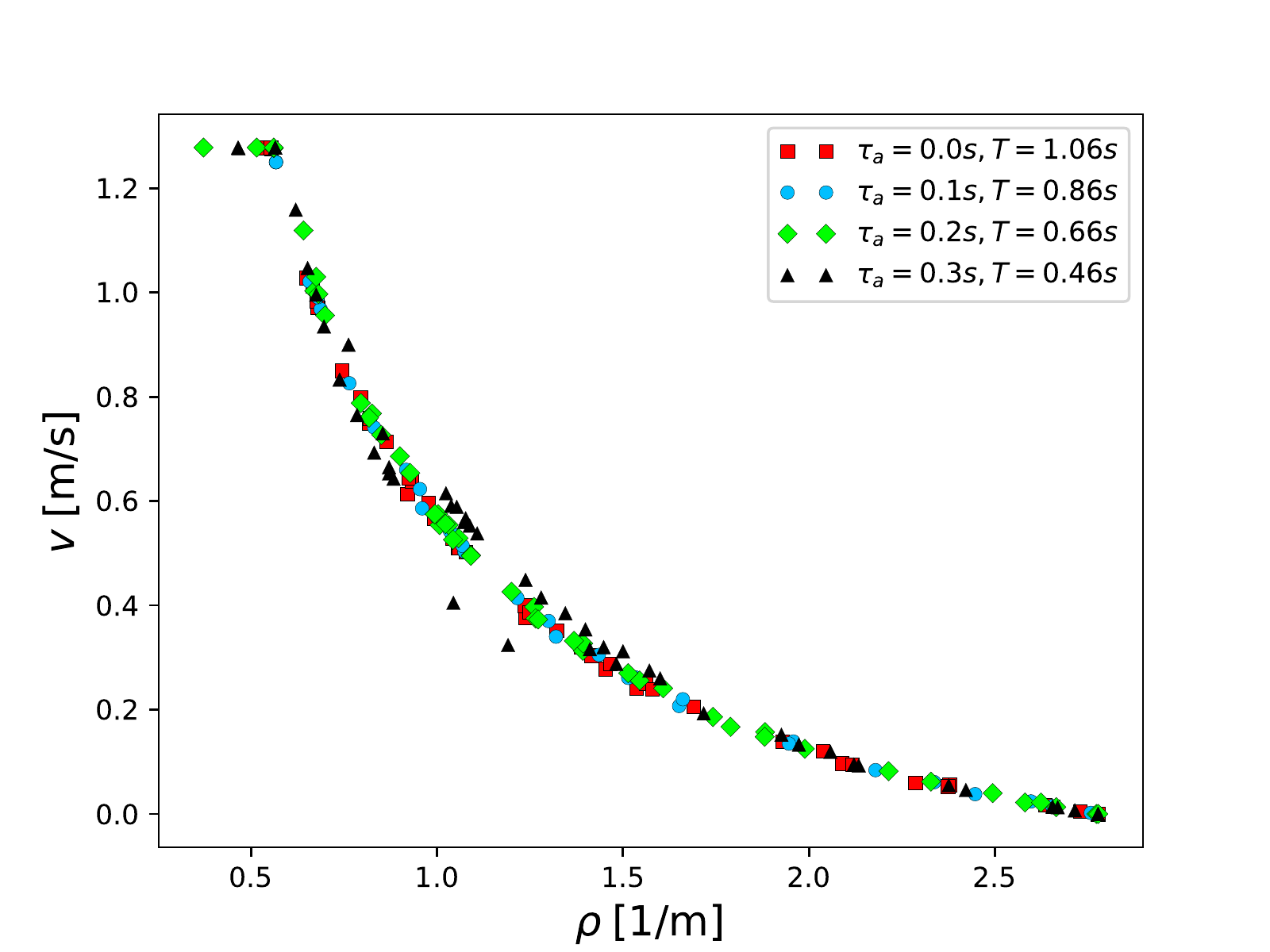}
    \caption{The fundamental diagram obtained in 1D simulations with $\tilde T = 1.06$.}
    \label{fig:Comparison_parameters}
\end{figure}

This becomes apparent after performing a basic stability analysis of the model.
Assuming a steady-state one-dimensional system, we can derive from the speed sub-model in Eq.~(\ref{equ:original velocity function}) the following relation
\begin{equation}
    V_{\rm{steady}}=\frac{1/\rho_{\rm{steady}}-2\cdot a_{\min}}{\tilde T},
\label{eq:sss}
\end{equation}
where $ V_{\rm{steady}}$ and $\rho_{\rm{steady}}$ are the speed and the density of pedestrians flow in steady state, and $\tilde T = T+2\tau_a$.
The parameter $\tau_a$ and the parameter $T$ in speed sub-model have the same influence on the dynamics. To confirm our assumption we perform numerically simulations by varying these two parameters, while maintaining a constant value of $\tilde T$. See figure~\ref{fig:Comparison_parameters}.

In the spirit of Occam's razor, we dispense with parameter $\tau_a$ and opt for a constant semi-axis $a_i$.

We can observe from figure~\ref{fig:Comparison_parameters} that although the values of $\tau_a$ are different in these simulations, the results obtained are almost identical when $\tilde T$ is constant.

The other semi-axis along the shoulder axis $b_i$ is defined according to~\cite{mohcine} as a linear function:
\begin{equation}
\label{equ:old_b}
     b_i=b_{\max}-(b_{\max}-b_{\min})\frac{V_i}{V_i^0},
\end{equation}
with $b_{\min}$ is the minimal semi-width when pedestrian $i$ reaches the desired velocity $V_i^0$ and $b_{\max}$ is the maximum semi-width reached when pedestrian $i$ is not moving \cite{mohcine}. 

We found in simulations with the CVM that this linear relationship does not provide satisfactory results.
Hence we introduce a new non-linear function inspired by the observation that pedestrians often reduce their occupied space in the vertical direction of motion by turning their body to walk faster and pass through narrow gaps that are smaller than the width of their shoulder. 

\begin{figure}[htbp]
    \centering
    \subfigure[$\gamma=0.1$]{\label{fig:sigmoid_beta}
    \includegraphics[width=0.45\linewidth]{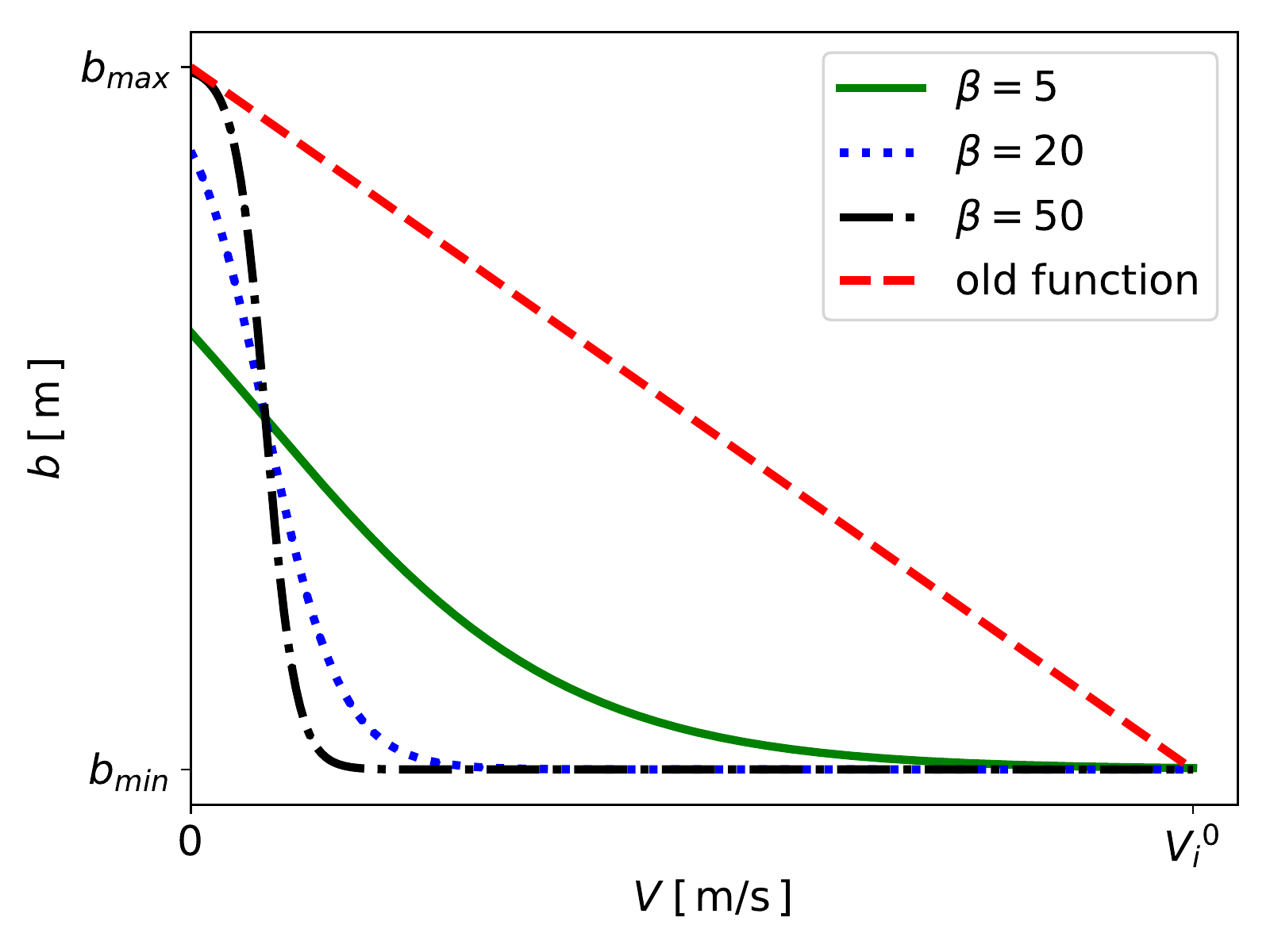}}
    \subfigure[$\beta=50$]{\label{fig:sigmoid_gamma}
    \includegraphics[width=0.45\linewidth]{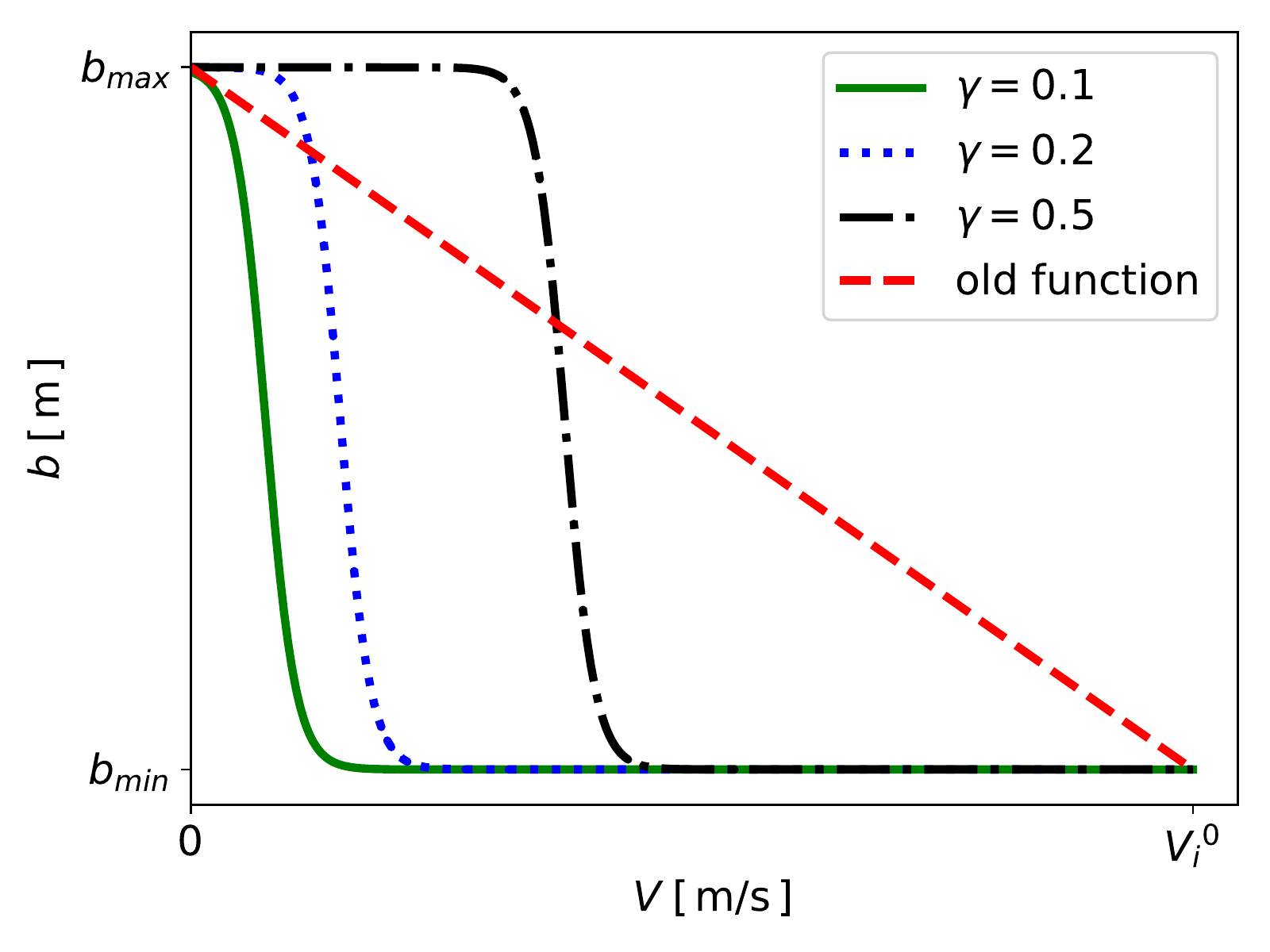}}
\caption{The new function of $b$ with different $\beta$ and $\gamma$, the left figure shows the curves of the function with same $\gamma=0.1$ but different $\beta$ while the right figure shows the curves of the function with same $\beta=50$ but different $\gamma$.}
\label{fig:new_b}
\end{figure}

We set 
\begin{equation}
\label{equ:new_b}
    b_i=b_{\min}+\frac{b_{\max}-b_{\min}}{1+e^{\beta \cdot(V_i-\gamma)}},
\end{equation}
which is a Sigmoid function, where the maximum semi-width $b_{\max}$ is equal to the half of a static pedestrian's width and $b_{\min}$ is equal to the half of a moving pedestrian's minimum width.
Parameters $\beta$ and $\gamma$ are used to adjust the shape of the function. 
Figure~\ref{fig:new_b} gives the curves of the function for different parameter values.

\begin{figure}[htbp]
    \centering
    \includegraphics[width=0.5\textwidth]{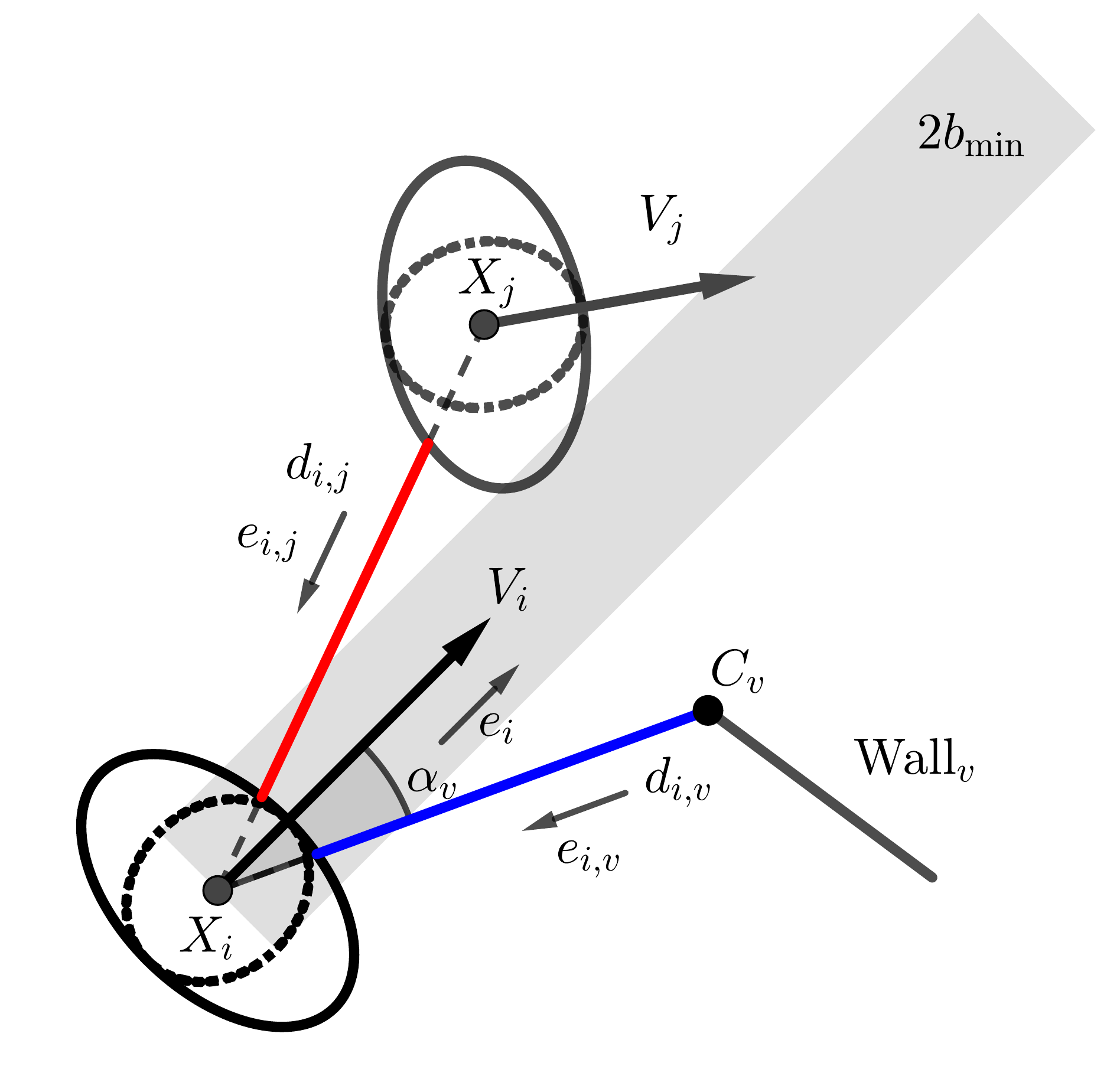}
    \caption{Notations used in the collision-free velocity model after extending the distance calculations between pedestrians from circle to velocity-based ellipse.
    $d_{i,j}$ is the length of red segment and $d_{i,v}$ is the length of blue segment.
    $X_i$ and $X_j$ are positions of pedestrians, $C_v$ is the closet point in wall $v$ to $X_i$, $V_i$ and $V_j$ are moving speeds of pedestrians, $\vec{e}_i$ is the moving direction of pedestrian $i$, $\vec{e}_{i,j}$ and $\vec{e}_{i,v}$ are the unit vectors from $X_j$ and $C_v$ to $X_i$, $\alpha_v$ is the angle between $\vec{e}_i$ and $-\vec{e}_{i,v}$.}
    \label{fig:EllipseCalculation}
\end{figure}

After defining the semi-axes of the ellipse, we extend the distance calculations from circle to velocity-based ellipse. 
See figure \ref{fig:EllipseCalculation}.
The ellipses in full line describe non-moving pedestrians, while the ellipses in the dashed line represent the pedestrians at the desired velocity.
$d_{i,j}$ is the distance between ellipses used to represent pedestrian $i$ and $j$, which is defined as the distance between the borders of ellipses $i$ and $j$, along a line connecting their centers. 
$d_{i,v}$ is the distance between the wall $v$ and pedestrian $i$, which is defined as the distance between the $C_v$ (the closest points in wall $v$ to the center of pedestrian $i$) and the border of ellipse used to present pedestrian $i$, along a line connecting the center of pedestrian $i$ and $C_v$. 

In the new equation of motion, the influence of the agents' shape is added as follows:
The moving direction $\vec{e}_i$ is calculated by Eq.~(\ref{equ:direction model with wall}), but the new definition of functions 
\begin{equation}
\label{equ:R and Rw}
    R(d_{i,j})=k\cdot \exp\Big(\frac{d_{i,j}}{D}\Big),\quad R_w(d_{i,v})=k_w\cdot \exp\Big(\frac{d_{i,v}}{D_w}\Big),
\end{equation}
are used. 
Then the velocity $V_i$ is obtained by
\begin{equation}
    V_i=\min\bigg\{V_{i}^0,\max\Big\{0,\frac{d_{i}}{T}\Big\},\max\Big\{0,\frac{dw_{i}}{T}\Big\}\bigg\},
\end{equation}
where
\begin{equation}
\label{equ:ellipse spacing in front i}
    d_i=\min_{j\in J_i}d_{i,j},\quad 
    dw_i=\min_{v\in JW_i} \frac{d_{i,v}}{\cos{\alpha_v}}.
\end{equation}
Here $J_i$ and $JW_i$ are the sets containing all pedestrians and walls in the direction of movement (i.e.\ the pedestrians and walls overlap with the grey area in figure \ref{fig:EllipseCalculation}). 
We set the width of the grey area to $2b_{\min}$ in the case a of velocity-based ellipse.
The comparison between the models describing agents with different shapes is given in section \ref{sec:simulation}.

\subsection{New direction sub-model}
After generalizing the model to ellipses, some unrealistic phenomena during simulation become visible.
First of all, backward movements occur very often, which is not realistic especially in evacuation scenarios. 
Second, an unnatural ``shaking'' appears during simulation, which is due to a strong fluctuation of the ellipse's orientation.

In the original model, the moving direction of pedestrian $i$ is calculated by combining individual desire direction $\vec{e}_i^{\ 0}$ and the neighbours' influence. 
Since the direction of neighbour's influence is from the center of pedestrians or closest point on the wall towards the center of the pedestrian $i$, the influence can be divided into two parts, one is the projection on $\vec{e}_i^{\ 0}$ and the other one is perpendicular to the projection part. 
The direction of the projection part is the reason for backward movements. 
Actually, pedestrians hardly choose a moving direction whose projection on $\vec{e}_i^{\ 0}$ is in the inverse direction of $\vec{e}_i^{\ 0}$. 
And the cause of the ``shaking'' is that pedestrian turn to $\vec{e}_i$ directly after calculation in the original model (0th order model). 

\begin{figure}[htb]
    \centering
    \includegraphics[width=0.5\textwidth]{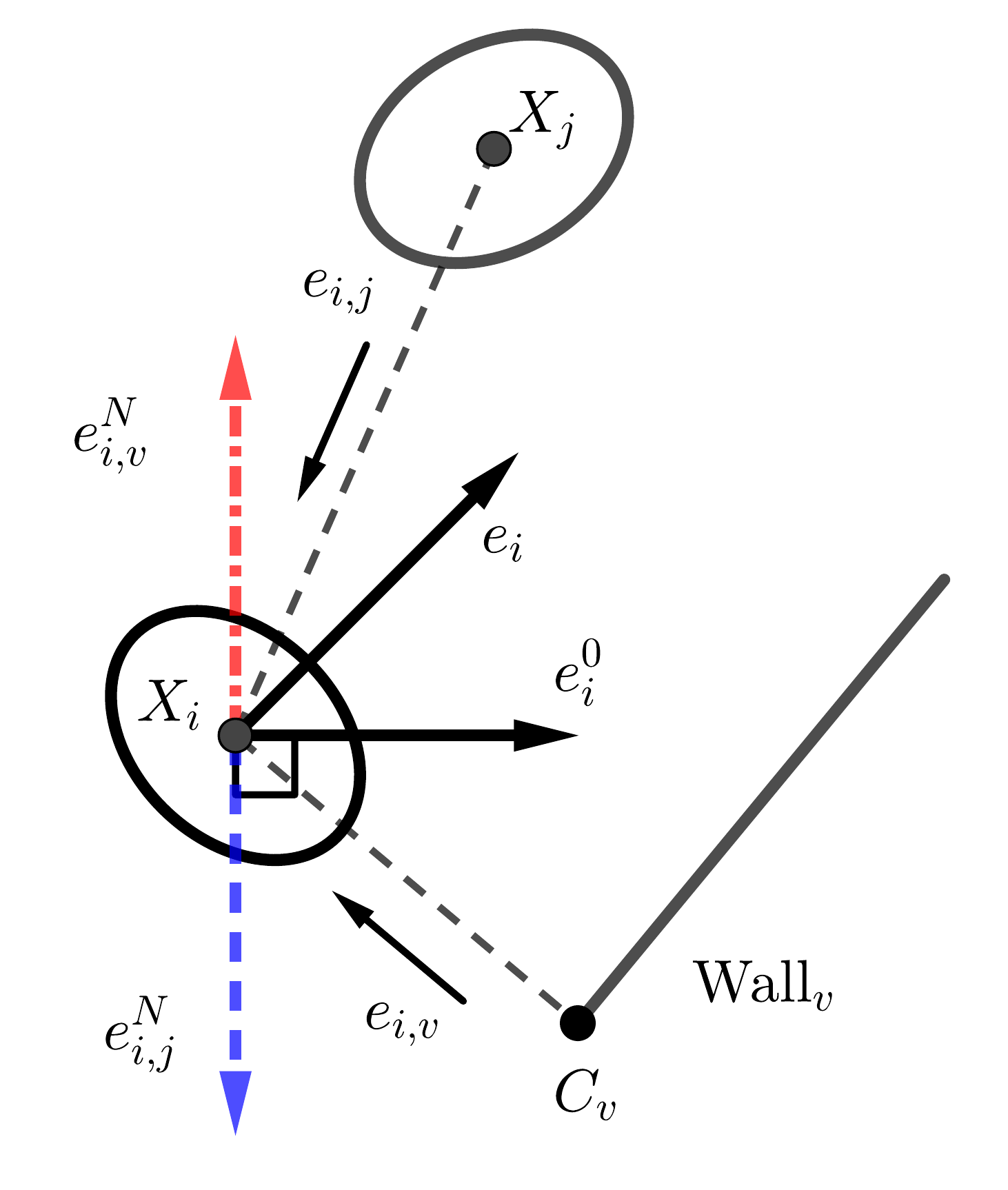}
    \caption{Notations used in the new direction sub-model.
    the direction of influence from pedestrian $j$ to pedestrian $i$ is vector represented by red chain line and the direction of influence from wall $v$ to pedestrian $i$ is vector represented by the blue dashed line.
    $X_i$ and $X_j$ are positions of the pedestrians, $C_v$ is the closet point in wall $v$ to $X_i$, $\vec{e}_i^{\ 0}$ and $\vec{e}_i$ are the desired moving direction and the actual moving direction of the pedestrian $i$, $\vec{e}_{i,j}$ and $\vec{e}_{i,v}$ are the unit vectors from $X_j$ and $C_v$ to $X_i$, $\vec{e}_{i,j}^{~N}$ and  $\vec{e}_{i,v}^{~N}$ are the directions used to calculate the influence of the pedestrians $j$ and walls $v$ act on pedestrian $i$ respectively.}
    \label{fig:NewDirection}
\end{figure}

Therefore, our solution has two parts, the projection of neighbours influence on $\vec{e}_i^{\ 0}$ is always equal to zero, and introducing a smoothing process (e.g.\ a relaxation process) in the direction sub-model. 
Based on this idea, we propose a new direction sub-model as shown in figure \ref{fig:NewDirection}, where $\vec{e}_i^{\ 0}$ is the desired moving direction of the pedestrian $i$ and $\vec{e}_i$ is the actual moving direction, $\vec{e}_{i,j}^{~N}$ and  $\vec{e}_{i,v}^{~N}$ are the new directions used to calculate the influence of the pedestrians $j$ and walls $v$ act on pedestrian $i$ respectively.

The new direction sub-model uses two steps to calculate the moving direction of a pedestrian. 
First, we use
\begin{equation}
\label{equ:newdirectioncalculation}
   \vec{E}_i=u_3 \cdot\bigg(\vec{e}_i^{\ 0}+\sum_{j\in N_i} R(d_{i,j})\cdot \vec{e}_{i,j}^{~N}+\sum_{v\in W_i} R_w(d_{i,v})\cdot \vec{e}_{i,v}^{~N}\bigg),
\end{equation}
to calculate the optimal moving direction of the pedestrian $i$, $u_3$ is a normalization constant such that $||\vec{E}_i||=1$. 
The repulsive function $R(d_{i,j})$ and $R_w(d_{i,v})$ are given in Eq.~(\ref{equ:R and Rw}) and the definition of $\vec{e}_{i,j}^{~N}$ and  $\vec{e}_{i,v}^{~N}$ are
\begin{equation}
\label{equ:new ped influence direction}
    \vec{e}_{i,j}^{~N}=\begin{cases}
    \vec{e}_i^{\ 0\bot}& \text{if }C_j>0, \\
    \vec{e}_i^{\ 0\bot}~\text{or}~-\vec{e}_i^{\ 0\bot}& \text{if }C_j=0, \\
    -\vec{e}_i^{\ 0\bot}& \text{if }C_j<0.
    \end{cases}\qquad 
     \vec{e}_{i,v}^{~N}=\begin{cases}
    \vec{e}_i^{\ 0\bot} & \text{if }C_v>0, \\
    \vec{e}_i^{\ 0\bot}~\text{or}~-\vec{e}_i^{\ 0\bot}& \text{if }C_v=0, \\
    -\vec{e}_i^{\ 0\bot}& \text{if }C_v<0.
    \end{cases}
\end{equation}
where 
\begin{equation}
C_j=\vec{e}_{i,j}\cdot\vec{e}_i^{\ 0\bot},\quad 
C_v=\vec{e}_{i,v}\cdot\vec{e}_i^{\ 0\bot}.
\end{equation}
Here, $\vec{e}_i^{\ 0\bot}$ is the vector obtained by rotating desired direction $\vec{e}_i^{\ 0}$ for $90^\circ$ counterclockwise.
According to Eq.~(\ref{equ:new ped influence direction}), influence from pedestrians and walls are decided not only by their position but also by the desired moving direction of the pedestrian $i$.
If the centers of pedestrians or the closest points in walls to the center of the pedestrian $i$ are located in the left area to $\vec{e}_i^{\ 0}$, the direction of influence is defined as right side perpendicular vector of $\vec{e}_i^{\ 0}$ and vice versa.
It should be noticed that there might be an extremely rare case when $C_j$ or $C_v$ is equal to zero.
In this case, the influence direction is decided by multiple factors, e.g. culture, gender. 
In order to simplify the model, the direction of influence is randomly chosen from $\vec{e}_i^{\ 0\bot}$ and $-\vec{e}_i^{\ 0\bot}$ in this case, corresponding to pedestrians avoiding front obstacles from the sides.
 
Then, we introduce a new relaxation time parameter $\tau$ in the direction sub-model, which is represented as
\begin{equation}
\label{equ:new moving direction}
\frac{d\vec{e}_i(t)}{dt} = \frac{\vec{E}_{i}(t)-\vec{e}_{i}(t)}{\tau},
\end{equation}
where $\vec{e}_i$ is the moving direction of the pedestrian $i$ and $\vec{E}_{i}$ is the optimal moving direction calculated via Eq.~(\ref{equ:newdirectioncalculation}). 
In this step, we change the direction sub-module from zero-order to first-order, which does not change the global first-order property of the original CVM.
By adjusting $\tau$, pedestrian can turn to moving direction smoothly.

\begin{figure}[ht]
    \centering
    \includegraphics[width=0.5\textwidth]{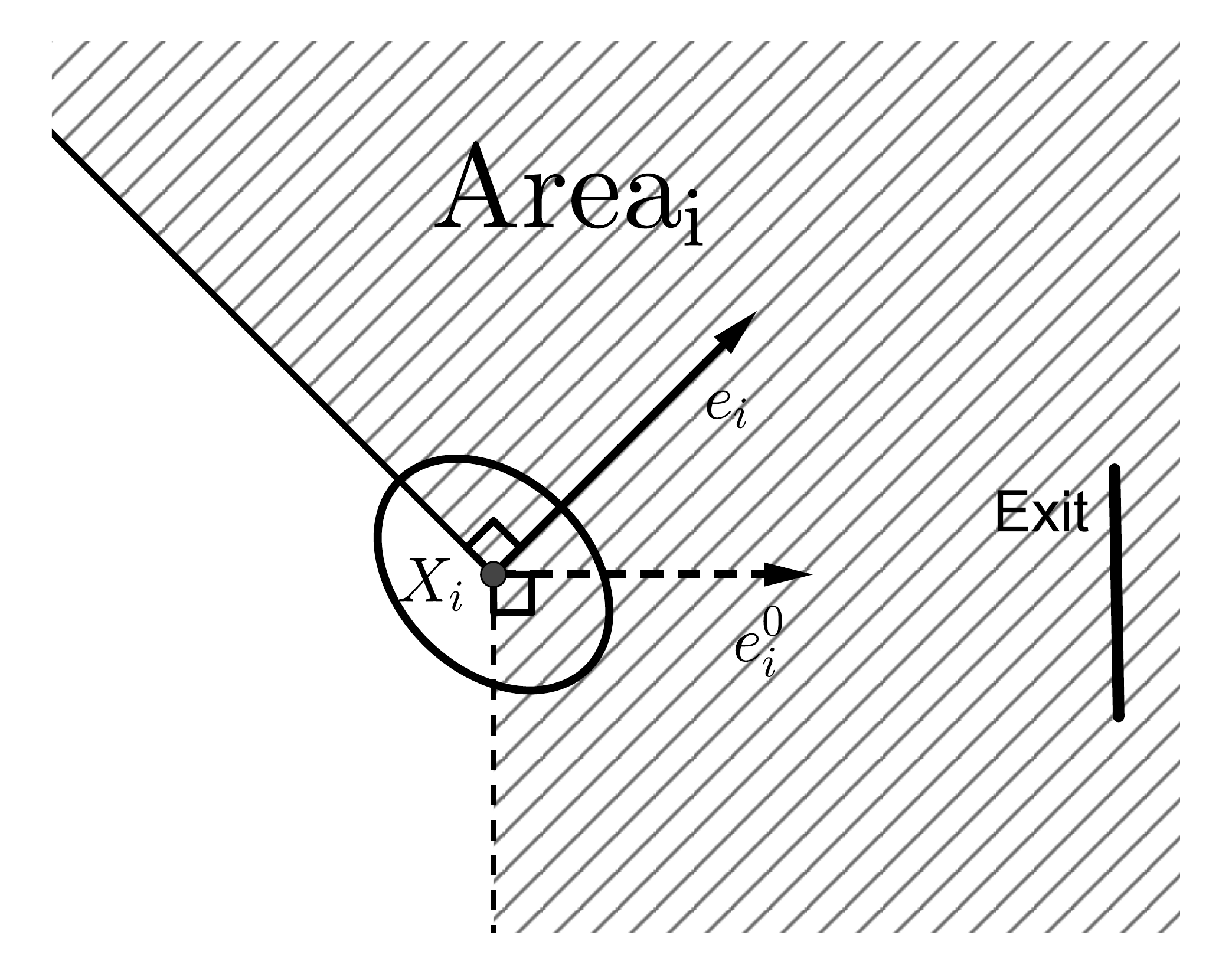}
    \caption{Dynamical vision area.
    $X_i$ is the position, $\rm{Area_i}$ is the hatching area, $\vec{e}_i^{\ 0}$ is the desired moving direction towards the exit, and $\vec{e}_i$ is the actual moving direction of pedestrian $i$.}
    \label{fig:visionarea}
\end{figure}

Besides, we use a dynamical vision area in this paper, which is the hatching area in figure \ref{fig:visionarea}. 
Only the pedestrians and walls located in ${\rm{Area}}_i$, which is the dynamical vision area the pedestrian $i$, influence the moving direction. 
The set contains all neighbours of the pedestrian $i$ in ${\rm{Area}}_i$ is
\begin{equation}
    N_i^{\rm{Area}}=\big\{j,\vec{e}_i \cdot \vec{e}_{i,j}<0\ \text{or}\  \vec{e}_i^{\ 0} \cdot \vec{e}_{i,j}<0\big\}.
\end{equation}
Here $\vec{e}_{i,j}$ is the vector from the center of neighbours towards the center of the pedestrian $i$. 
As for the walls, only when two vertices of a wall are both in ${\rm{Area}}_i$, this wall influences the moving direction of the pedestrian $i$.
Vision area of the pedestrian $i$ is decided by his desired moving direction $\vec{e}_i^{\ 0}$ and his actual moving direction $\vec{e}_i$. 
This means a pedestrian choose the best moving direction according to the neighbours and walls located in the half area in front of his moving direction and the half area in front of his desire moving direction. 
The dynamical vision area is based on the idea that pedestrians will turn their heads to obtain the environmental information of the areas in front of their desired moving directions if their actual moving directions deviate from the desired moving directions.
Using this dynamical vision area can eliminate some unrealistic block occurred between agents when using fixed vision area in the simulation.

\begin{figure}[htbp]
    \centering
    \includegraphics[width=0.7\textwidth]{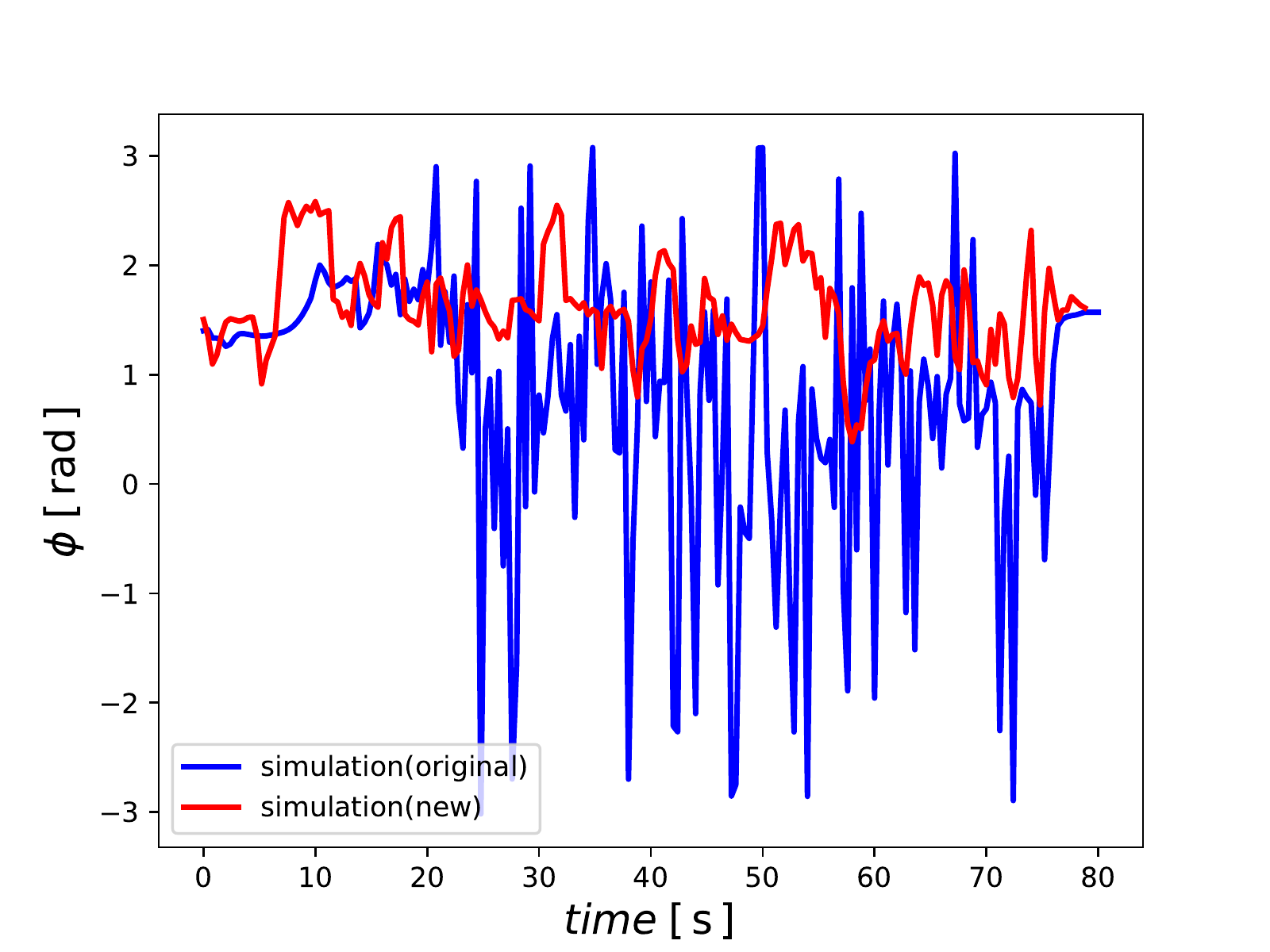}
    \caption{
    Time serie of $\phi$.
    $\phi$ is the angle between the the moving direction of a pedestrian and the x-axis.
    }
    \label{fig:unnatural_shaking}
\end{figure}

These enhancements can almost eliminate the phenomena of backward movement and ``shaking'' in the simulation, as shown in figure \ref{fig:unnatural_shaking}.

$\phi$ is defined as the angle between the the moving direction of a pedestrian and the x-axis.. 
As we can see in figure \ref{fig:unnatural_shaking}, the blue line  (original model) shows strong fluctuation of the angle over time compared with the red line (our extension).

In the next section we further show a systematic comparison of both models.
  
\section{Simulation results}
\label{sec:simulation}
In this section, the comparison and analysis of models with different shapes and different direction sub-models are given.
The simulations in this section are executed with Euler scheme using a time step $\Delta t=0.05~s$. 
The update of the pedestrians is parallel in each step.

First, we perform simulations in a 26~m corridor with periodic boundary condition and measure the 1D fundamental diagram in a two meters long area located in the middle of the corridor. 
The shape of agents in these simulations, as well as the direction sub-model, are insignificant for the outcome of the simulation since pedestrians can not overtake others walking in front. 
Hence, we can focus on the validation of the speed sub-model and the relation between the velocity and the required spacing in front.
 
The values of parameters are shown in table \ref{tab:parameters of 1D}.
The desire speed of the pedestrians is 1.34~m/s. 
The shape of agents is circular with a constant radius $a$. 
The value of $a$ is 0.18~m and the value of $T$ is 1.06~s, which are obtained from the linear relationship of required length and velocity \cite{aminT}.

\begin{table}[ht]
    \centering
    \setlength{\belowcaptionskip}{5pt}
    \caption{Parameters of CVM in one-dimensional scenario}
    \begin{tabular}{|c|c|c|c|c|c|c|c|c|}
    \hline
    \ & $V^0~\rm(m/s)$& $a~\rm(m)$& $T~\rm(s)$& $k$& $D~\rm(m)$& $k_w$& $D_w~\rm(m)$\\
    \hline
    1D& 1.34& 0.18& 1.06& 3.0& 0.1& 6.0& 0.05\\
    \hline
    \end{tabular}
    \label{tab:parameters of 1D}
\end{table}

\begin{figure}[htbp]
    \centering
    \includegraphics[width=0.7\textwidth]{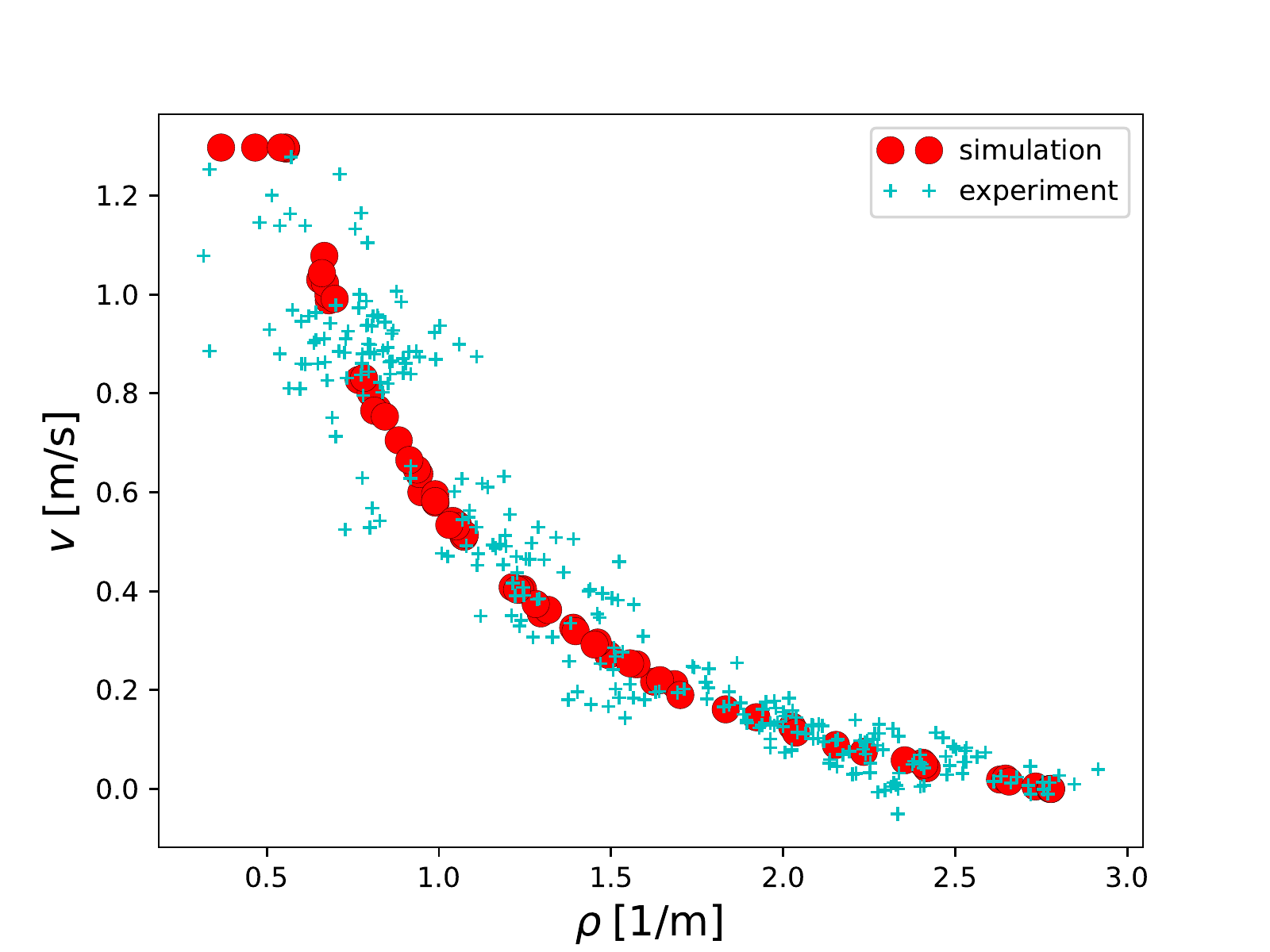}
    \caption{Velocity-density relation (fundamental diagram) in one-dimensional scenario, compared with experimental data \cite{1DdataBottleneck}.}
    \label{fig:1D}
\end{figure}

The simulation results in the 1D case are shown in figure \ref{fig:1D}.
We realize that the obtained 1D fundamental diagram fit well with the experimental data.

In the second step, we investigate the effect of the agent's shape on the two-dimensional fundamental diagram. 
The simulation scenario is a 26$\times$1.8~m$^2$ corridor with periodic boundary conditions.
We measure the 2D fundamental diagram of models which describing agent with different shapes. 
We use three kinds of shapes here, circles with constant radius, ellipses with constant $a$ and variable $b$ as defined in Eq.~(\ref{equ:old_b}) and ellipses with constant $a$ and variable $b$ as defined in Eq.~(\ref{equ:new_b}). 

The value of $V^0$, $a$, $T$ and parameters in direction sub-model are the same as in the one-dimensional case. 
Table \ref{tab:parameters of 2D} summarizes the value of other parameters.

\begin{table}[ht]
    \centering
    \setlength{\belowcaptionskip}{5pt}
    \caption{Parameters of CVM in two-dimensional scenario}
    \begin{tabular}{|c|c|c|c|c|c|}
    \hline
    \ & $b_{\min}~\rm(m)$& $b_{\max}~\rm(m)$& $b$ function& $\beta$& $\gamma$\\
    \hline
    constant circle& $\setminus$& $\setminus$& $\setminus$& $\setminus$& $\setminus$\\
    \hline
    original ellipse&  0.15& 0.25& (\ref{equ:old_b})& $\setminus$& $\setminus$\\
    \hline
    new ellipse& 0.15& 0.25& (\ref{equ:new_b})& 50& 0.1\\
    \hline
    \end{tabular}
    \label{tab:parameters of 2D}
\end{table}

\begin{figure}[htbp]
    \centering
    \includegraphics[width=0.7\textwidth]{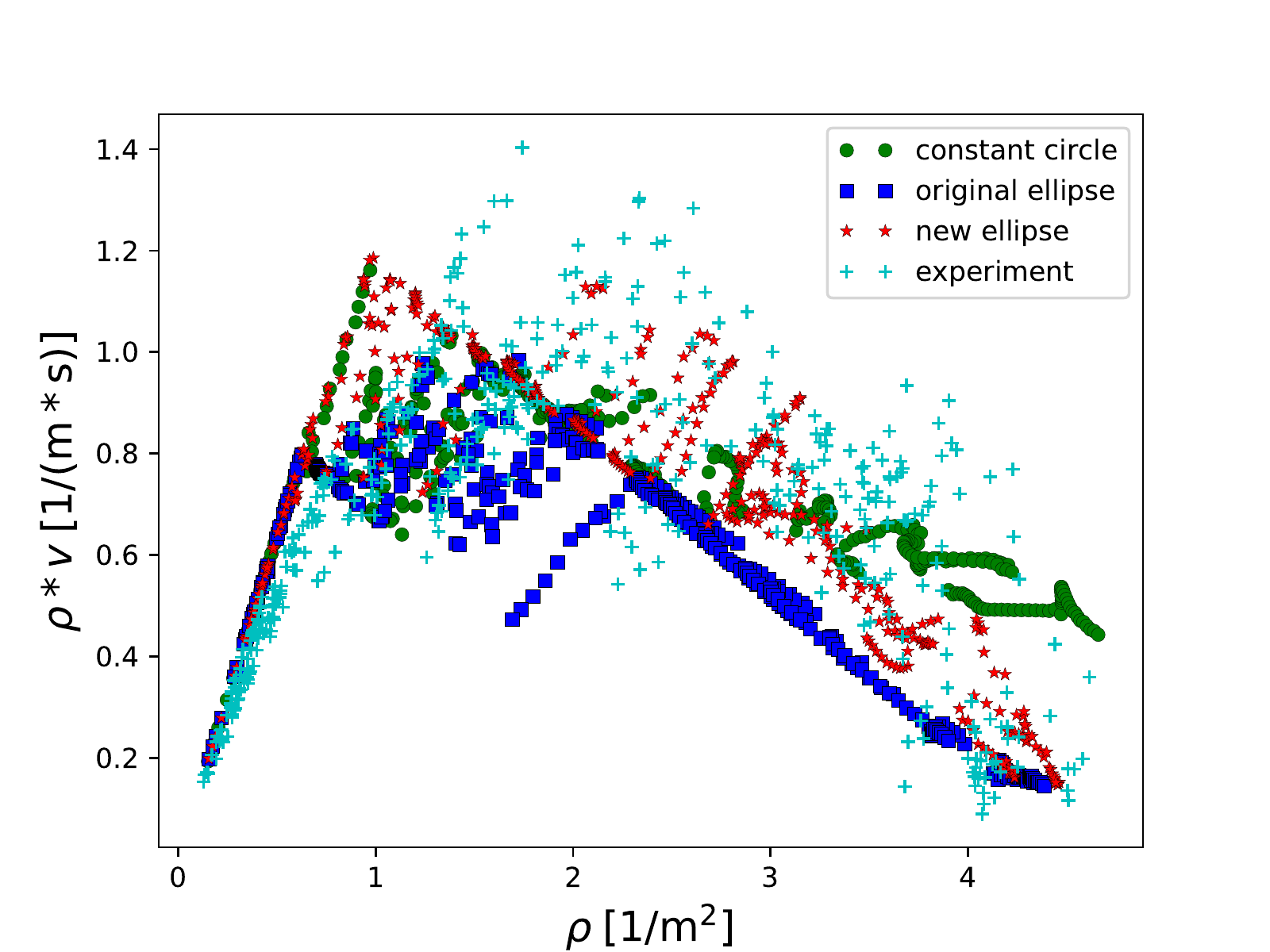}
    \caption{Relation between flow and density (fundamental diagram) in two-dimensional scenario, compared with experimental data obtained in the Hermes-project \cite{2Ddata}.}
    \label{fig:2D}
\end{figure}

The simulation results of 2D case are shown in figure \ref{fig:2D}.
From figure \ref{fig:2D}, we can get the result that the shape of agents in the model influence the fundamental diagram in the two-dimensional scenario, especially in high-density area. 
The results obtained with constant circle and ellipse with variable $b$ defined as Eq.~(\ref{equ:old_b}) both have deviation with experimental data in high-density area while using ellipse with variable $b$ defined as Eq.~(\ref{equ:new_b}) can obtain 2D fundamental diagram which is closer to the experimental results. 
That means the new function for $b$ we proposed has a positive impact on the simulation result.

Then, we perform simulations in bottleneck scenarios \cite{1DdataBottleneck}. 
We measure the relation between the flow in the middle of the bottleneck and the width of the bottleneck which is adjusted from 1.0~m to 2.5~m in our simulations.
As we mentioned before, we can observe some unusual behaviour during the simulation. 
Besides, we observe that the distribution of the pedestrians in front the of bottleneck is different from the experiment. 
The new direction sub-model proposed in the previous section can eliminate these unusual phenomena. 

\begin{figure}[ht]
    \centering
    \includegraphics[width=0.7\textwidth]{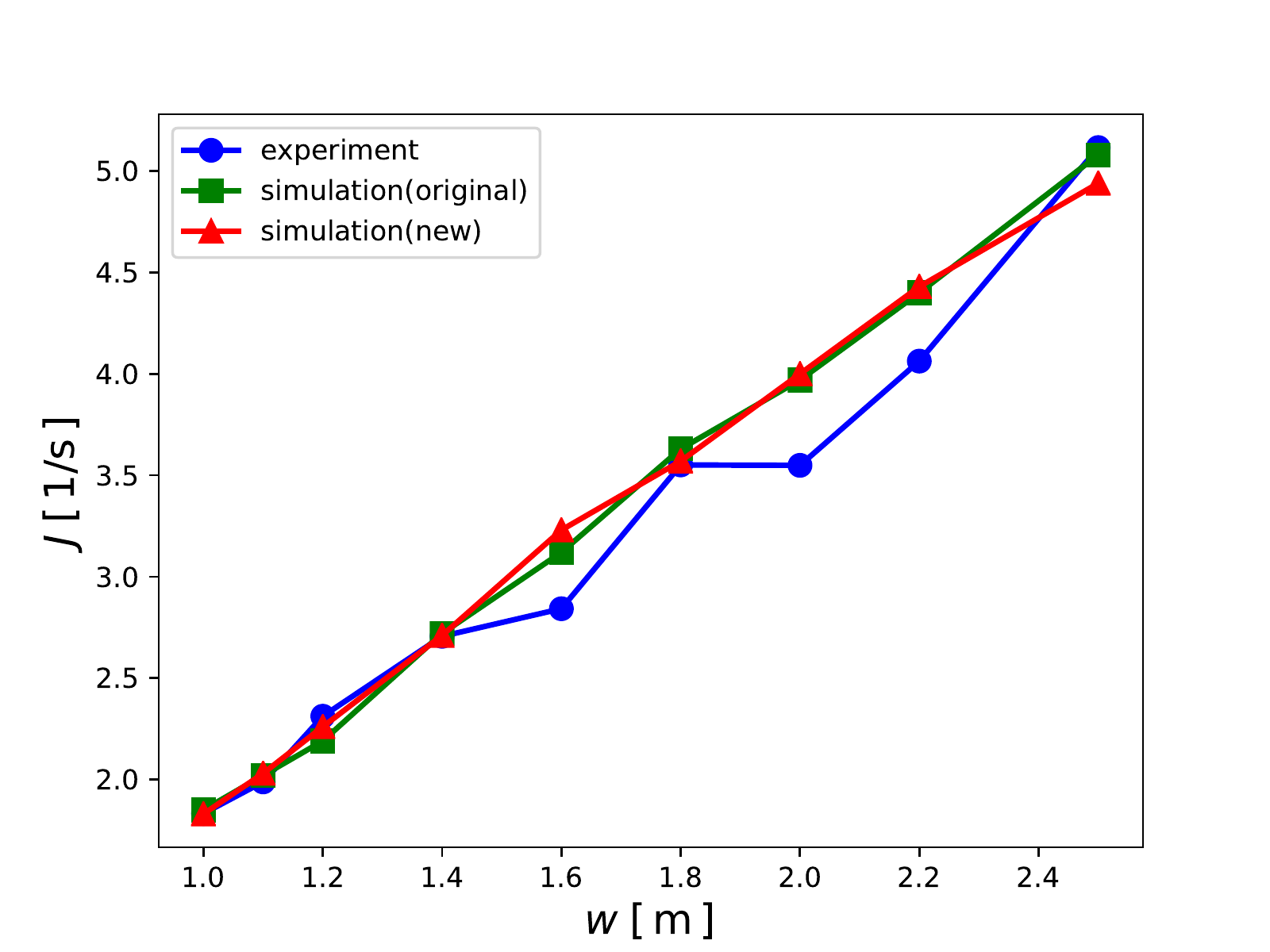}
    \caption{Relation between the flow in the middle of the bottleneck and the width of bottleneck, compared with experimental data[23].}
    \label{fig:bottleneck}
\end{figure}

In order to compare the simulation results of original and new direction sub-model fairly, we adjust the value of the parameter $T$ to make the flow-width relation obtained from the simulation results as close to the relation obtained from experimental data as possible. 
The shape of the pedestrian in original and new model are both the new dynamical ellipse we proposed in previous section, the value of $a$, $b_{\min}$, $b_{\max}$, $\beta$ and $\gamma$ are given in table \ref{tab:parameters of 2D}, the value of $k$, $D$, $k_w$, $D_w$ are provided in table \ref{tab:parameters of 1D}. 
The desired speeds of the pedestrians are Gaussian distributed with a mean of 1.34~m/s and a standard deviation of 0.26~m/s \cite{Weidmann}.
After validation, the value of $T$ in the original model is 0.5~s and in the new model is 0.45~s.
The value of new parameter $\tau$ introduced in new direction sub-model is 0.3~s. 
The relations obtained are shown in figure \ref{fig:bottleneck} and compared with experimental data. 
In figure \ref{fig:bottleneck} we can find the relation obtained from simulation results of the original and new model both very close to the experimental data.

Since the purpose of our extension is to eliminate backward movement and shaking phenomenon. 
We compare two indexes to prove that our extensions are useful.
The first one is the backward movement proportion
\begin{equation}
    O=\frac{\displaystyle\sum_{i=1}^{N} \sum_{k=0}^{M_i} O_i(k\cdot\Delta t)}{\displaystyle\sum_{i=1}^{N} M_i },
\end{equation}
where $\Delta t$ is the time step size in the simulation, $M_i*\Delta t$ is the simulation duration of pedestrian $i$, $N$ is the number of pedestrians in the simulation and 
\begin{equation}
    O_i(t)=
    \begin{cases}
    1, &\vec{e}_i(t)\cdot \vec{e}_i^{\ 0}(t)<0\\
    0, &\rm{else}
    \end{cases},
\end{equation}
where $\vec{e}_i(t)$ is the moving direction of pedestrian $i$. 
This definition means that when the angle between the actual moving direction and the desired moving direction of a pedestrian is greater than 90 degrees, we regard it as a backward movement.

\begin{figure}[htbp]
    \centering
    \includegraphics[width=0.7\textwidth]{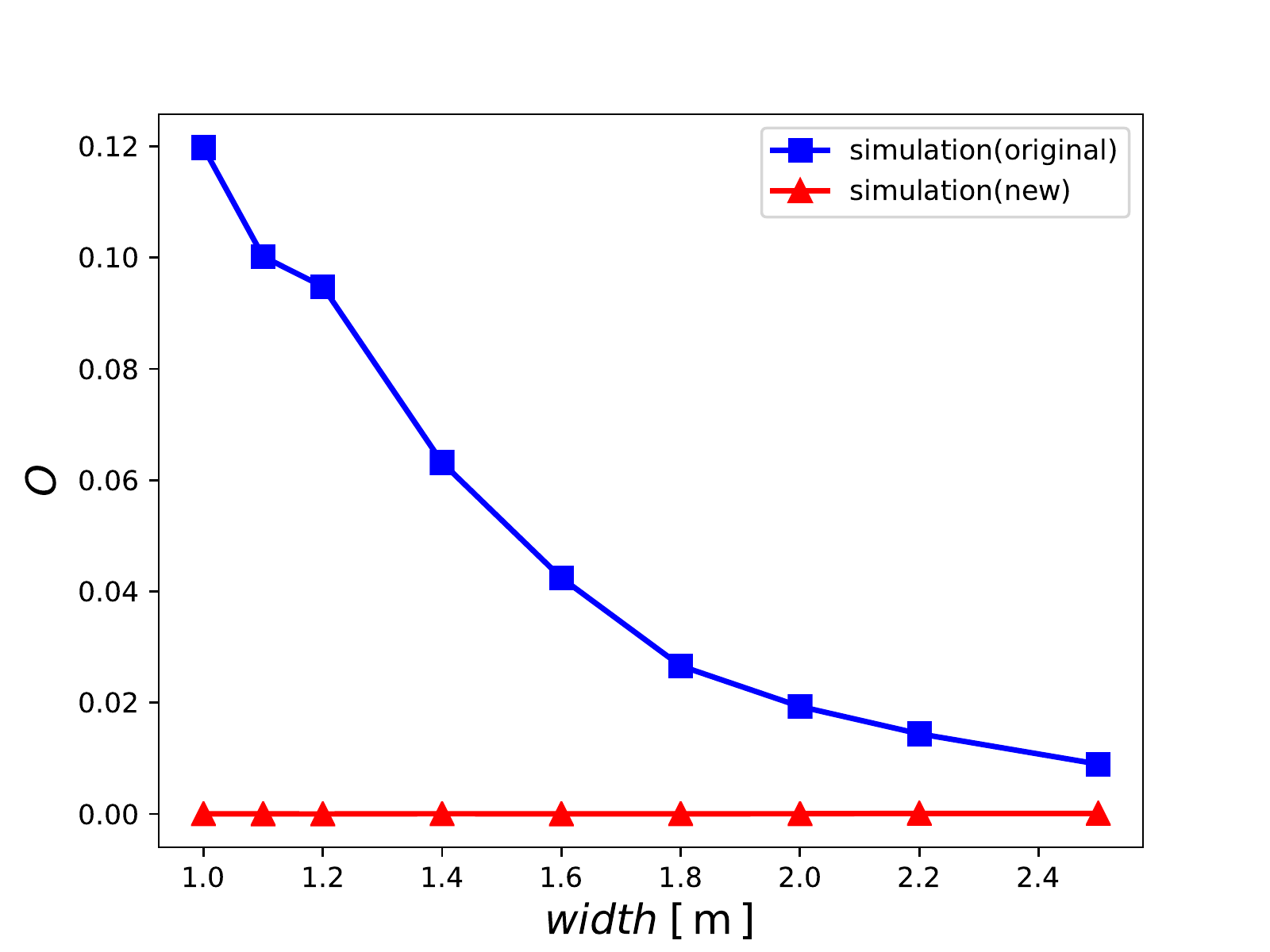}
    \caption{Proportion of backward movement in bottleneck scenarios with different widths from 1.0~m to 2.5~m.}
    \label{fig:backward}
\end{figure}

We calculate the proportion of backward movement from the simulation results of the original model and new model in bottleneck scenarios with different widths from 1.0~m to 2.5~m.
The results are shown and compared in figure \ref{fig:backward}.

From figure \ref{fig:backward}, we can find that the proportion of backward movement significantly decrease in the new model compared to the original model. 
Therefore our extension eliminate the unrealistic backward movement.

\begin{figure}[htbp]
    \centering
    \includegraphics[width=0.7\textwidth]{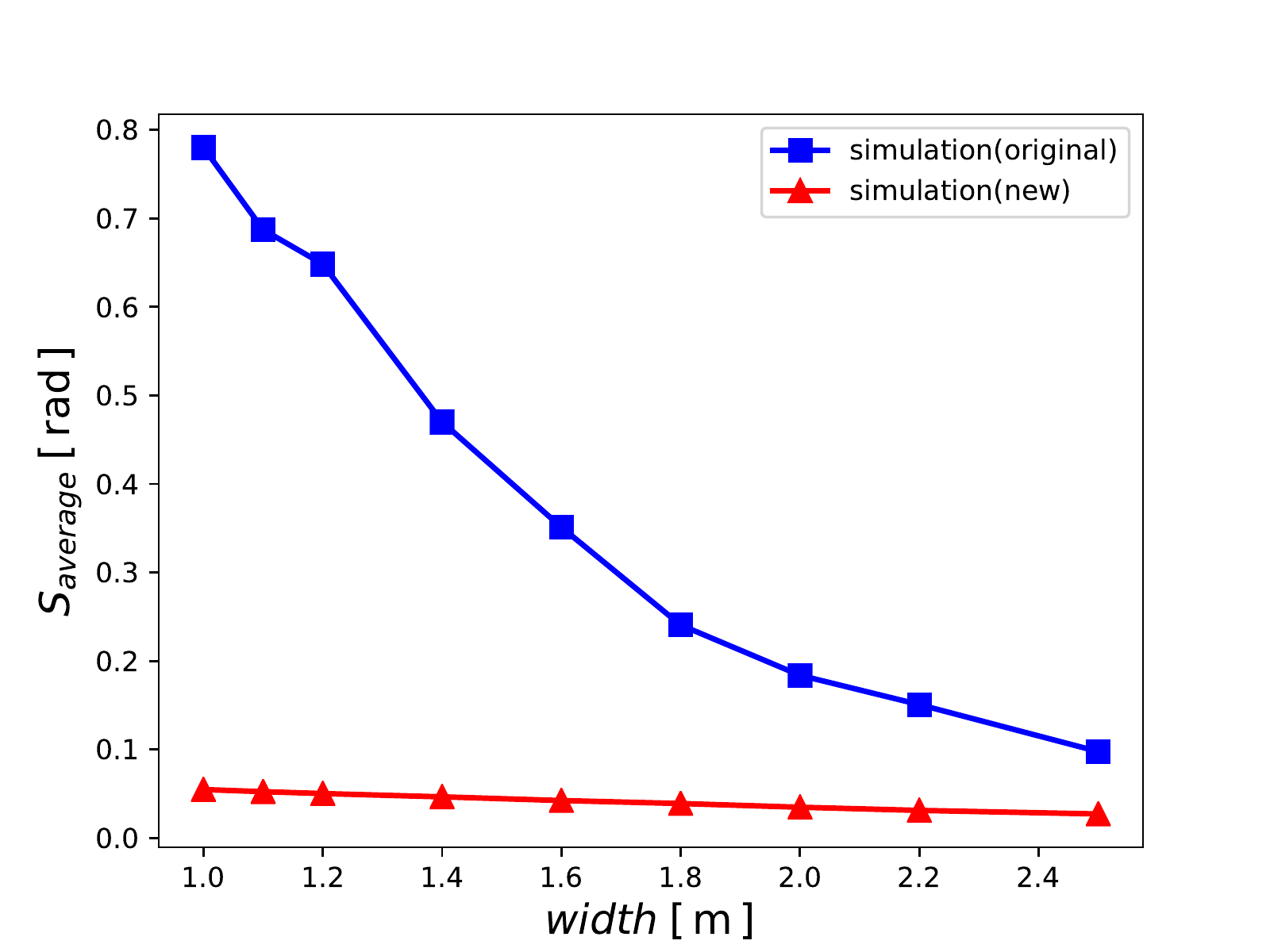}
    \caption{Average angular variation in moving direction per pedestrian per frame in bottleneck scenarios with different widths from 1.0~m to 2.5~m.}
    \label{fig:shaking}
\end{figure}

The second index is the average angular variation in moving direction per pedestrian per frame, which is presented as
\begin{equation}
    S_{average}=\frac{\displaystyle\sum_{i=1}^{N} \sum_{k=1}^{M_i} S_i(k\cdot\Delta t)}{\displaystyle\sum_{i=1}^{N} (M_i-1) }
\end{equation}
with
\begin{equation}
    S_i(t)=\left|\angle\big[\vec{e}_i(t),\vec{e}_i(t-\Delta t)\big] \right|,
\end{equation}
where the definition of $\angle\big[\vec{e}_i(t),\vec{e}_i(t-\Delta t)\big]$ is the angle between $\vec{e}_i(t)$ and $\vec{e}_i(t-\Delta t)$. The definition of moving direction $\vec{e}_i(t)$ is the same as before.
$S_i(t)$ is the absolute value of the angle between moving direction in the current time step and the previous one. 
We compare this index for the new model and the original model.
The results are presented in figure \ref{fig:shaking}.

It can be observed in figure \ref{fig:shaking} that in the new model the pedestrians change less their direction than the pedestrians in the original model, which is in line with the fact that pedestrians prefer to keep their direction instead of changing it. 
Compared within the original model, agents no longer shake frequently.

\begin{figure}[htbp]
    \centering
    \subfigure[density, experiment]{\label{fig:subfig:a}
    \includegraphics[width=0.3\linewidth]{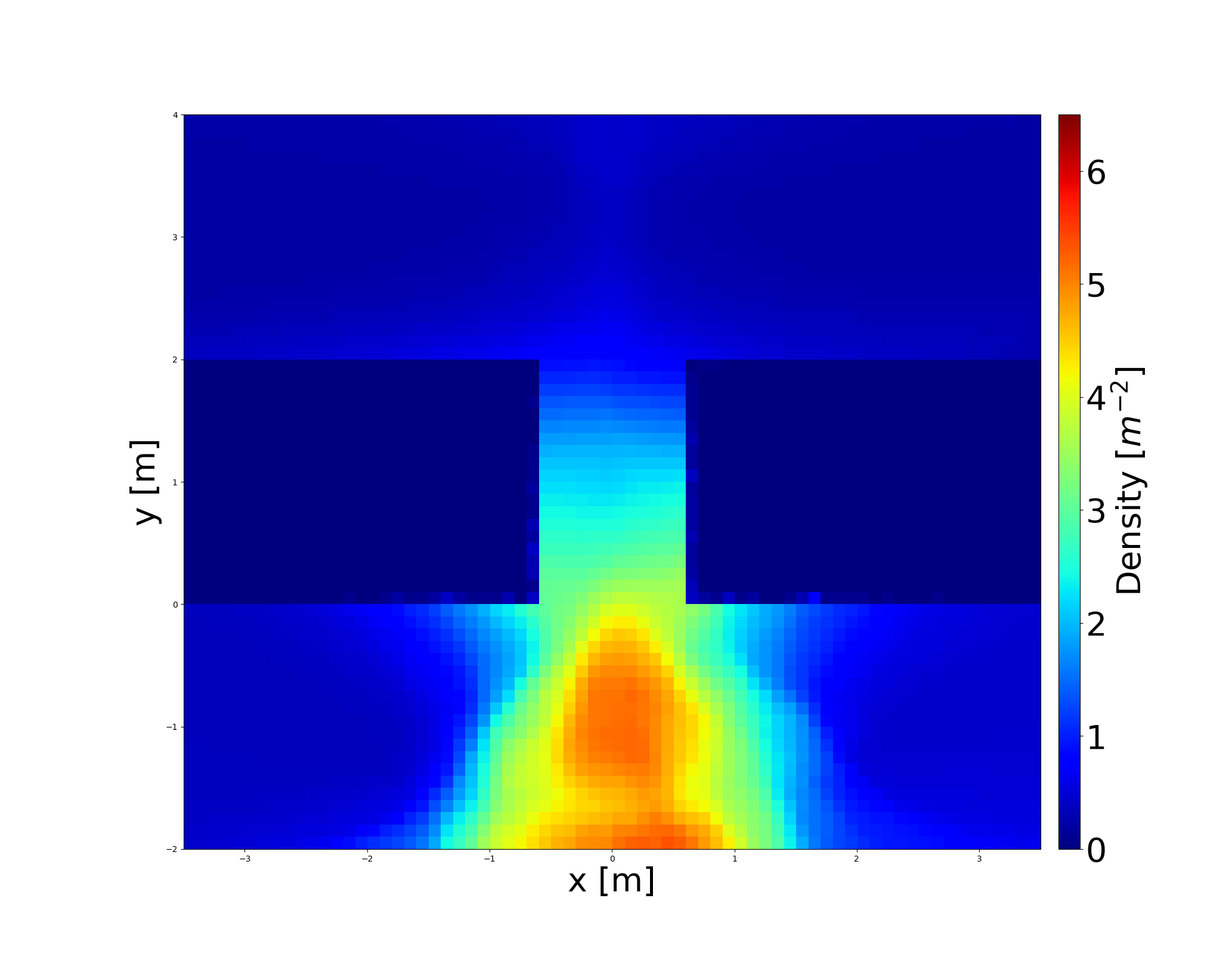}}
    \subfigure[density, original model]{\label{fig:subfig:b}
    \includegraphics[width=0.3\linewidth]{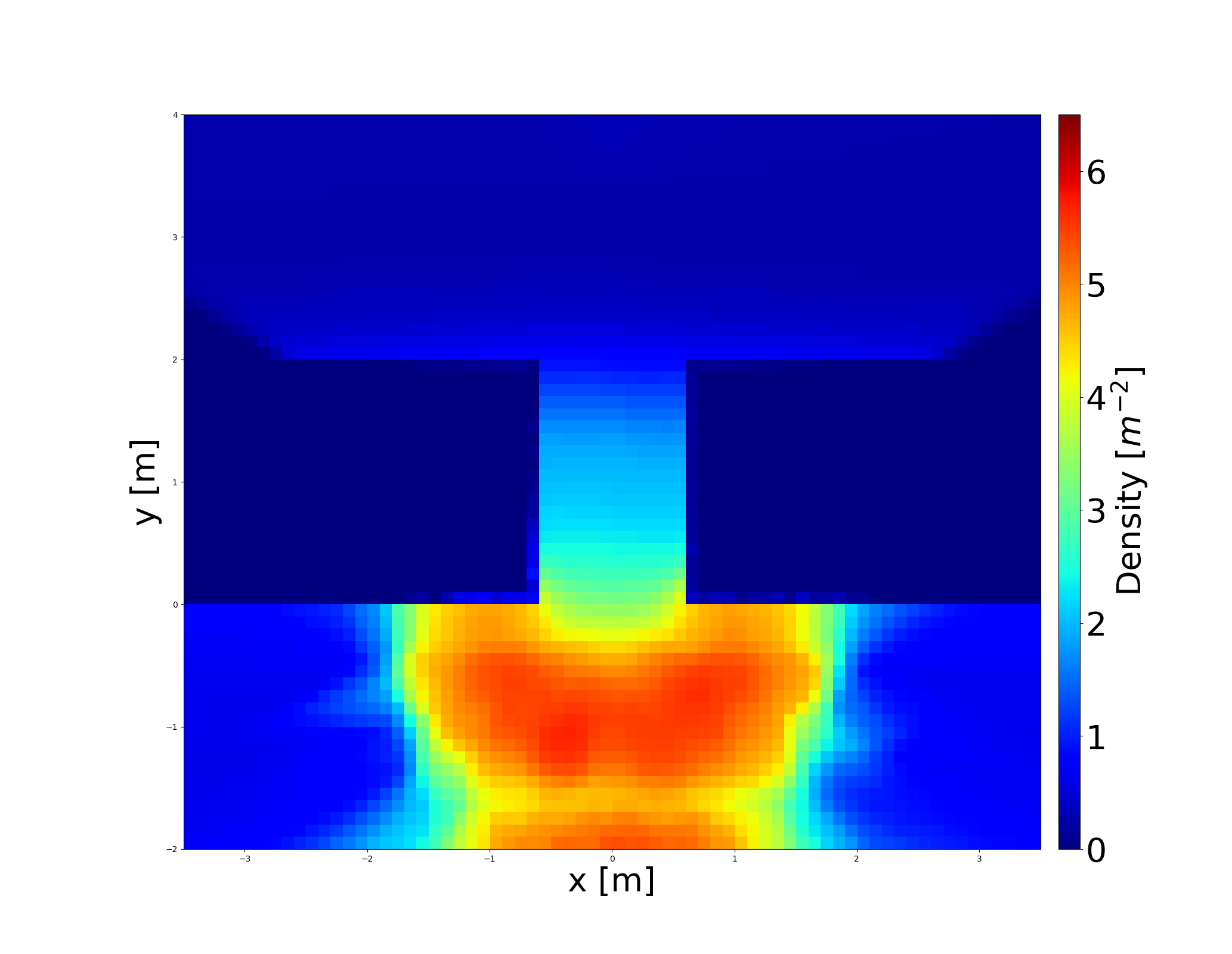}}
    \subfigure[density, new model]{\label{fig:subfig:c}
    \includegraphics[width=0.3\linewidth]{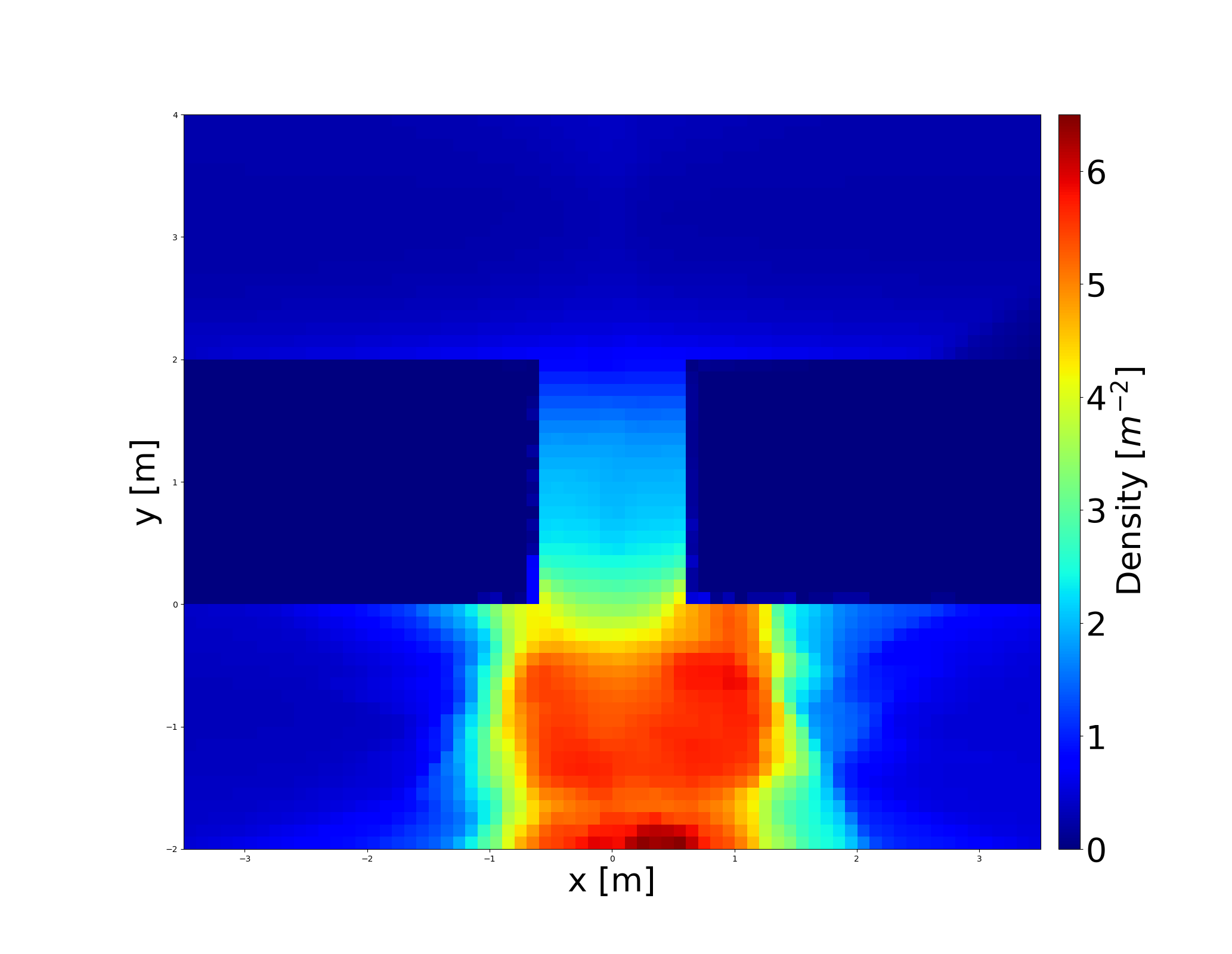}}
    \vfill
    \subfigure[velocity, experiment]{\label{fig:subfig:d}
    \includegraphics[width=0.3\linewidth]{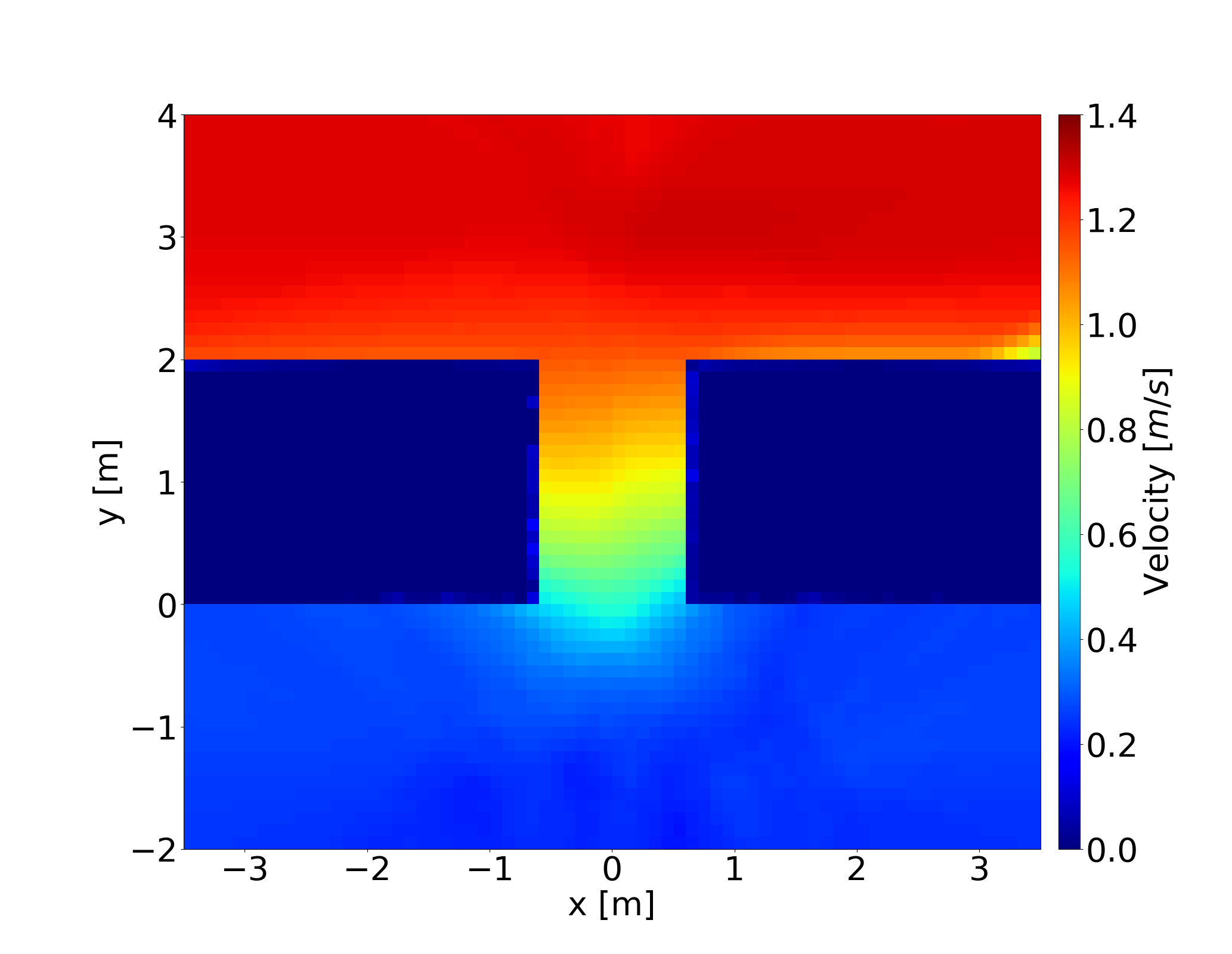}}
    \subfigure[velocity, original model]{\label{fig:subfig:e}
    \includegraphics[width=0.3\linewidth]{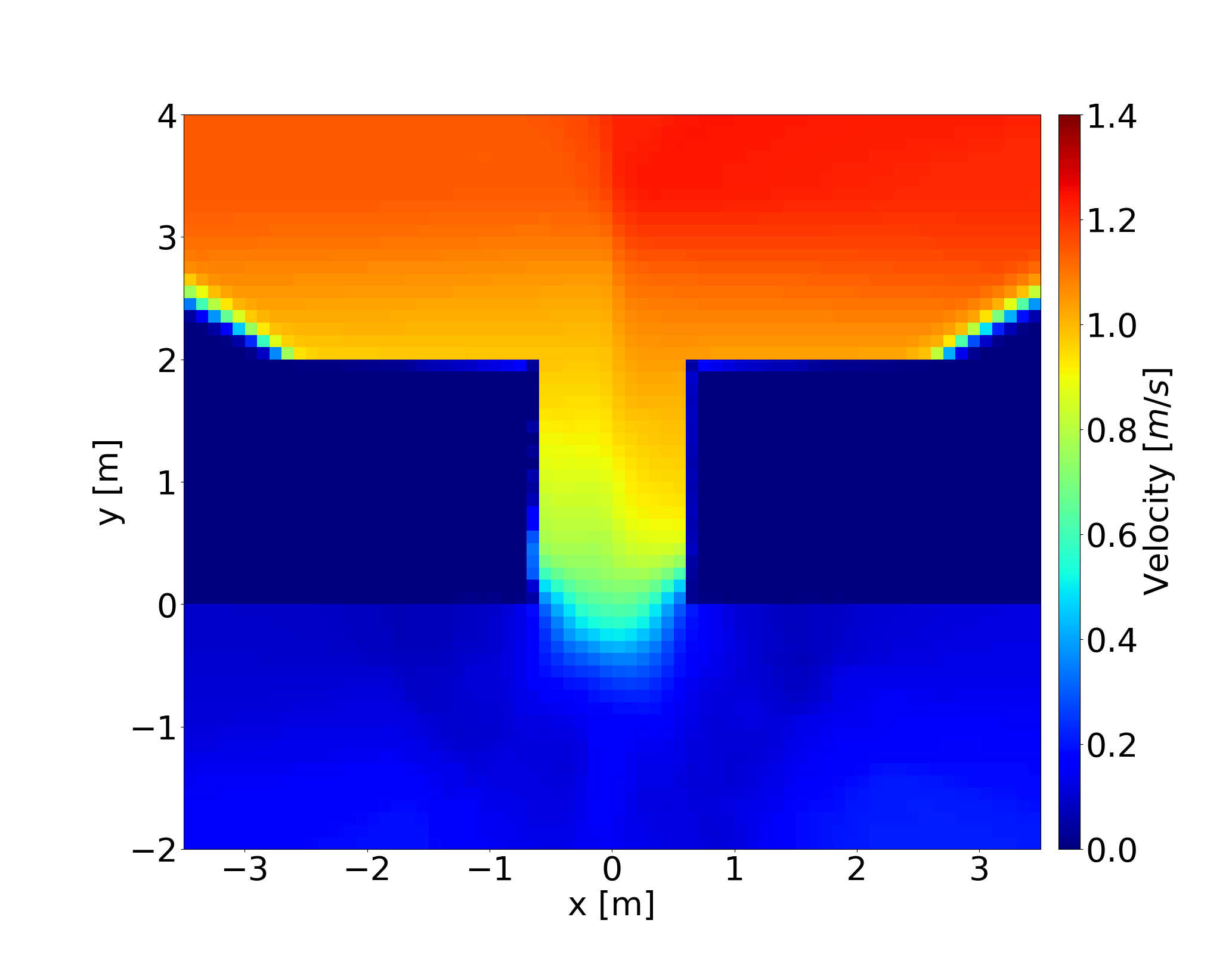}}
    \subfigure[velocity, new model]{\label{fig:subfig:f}
    \includegraphics[width=0.3\linewidth]{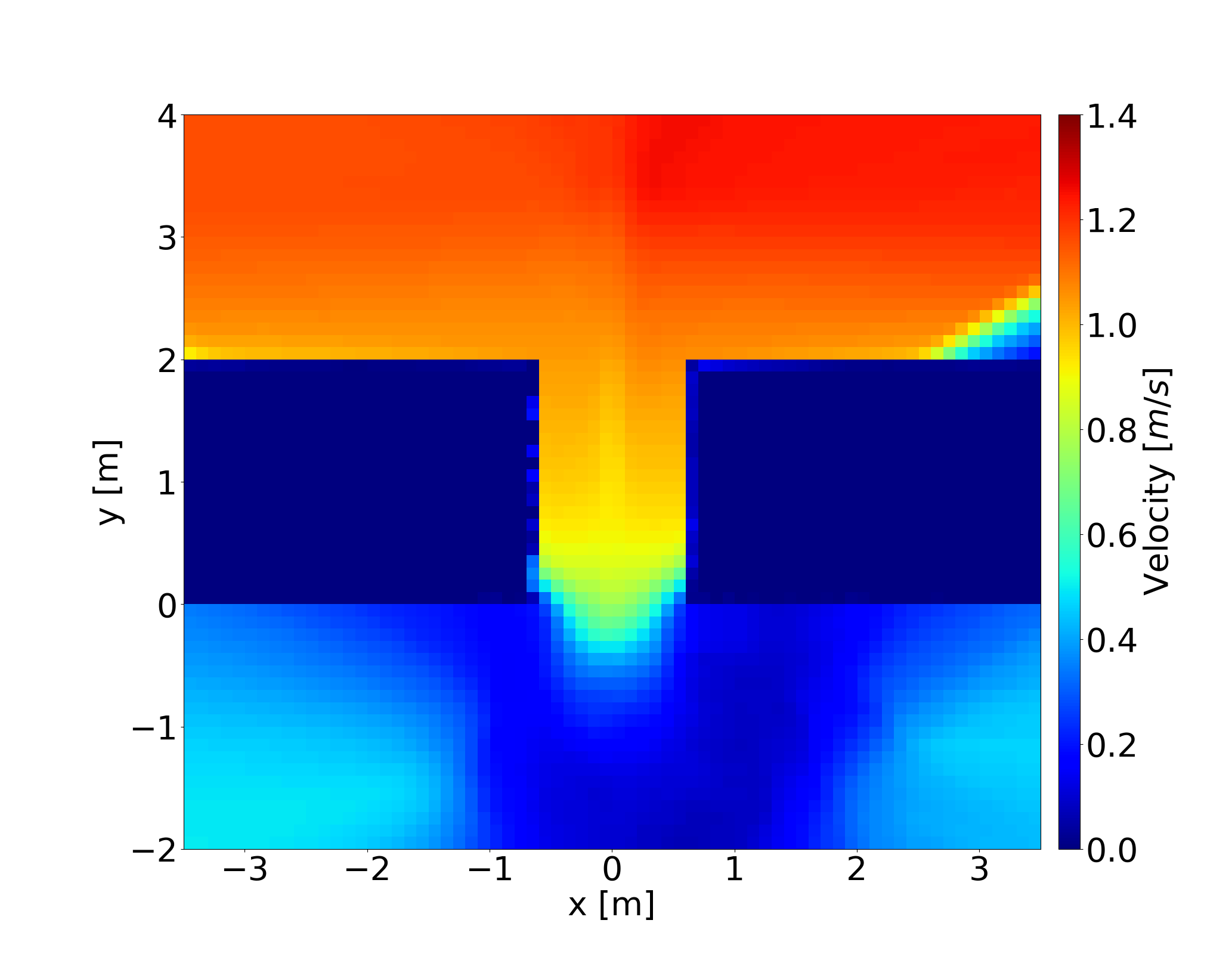}}
    \vfill
    \subfigure[Specific flow, experiment]{\label{fig:subfig:g}
    \includegraphics[width=0.3\linewidth]{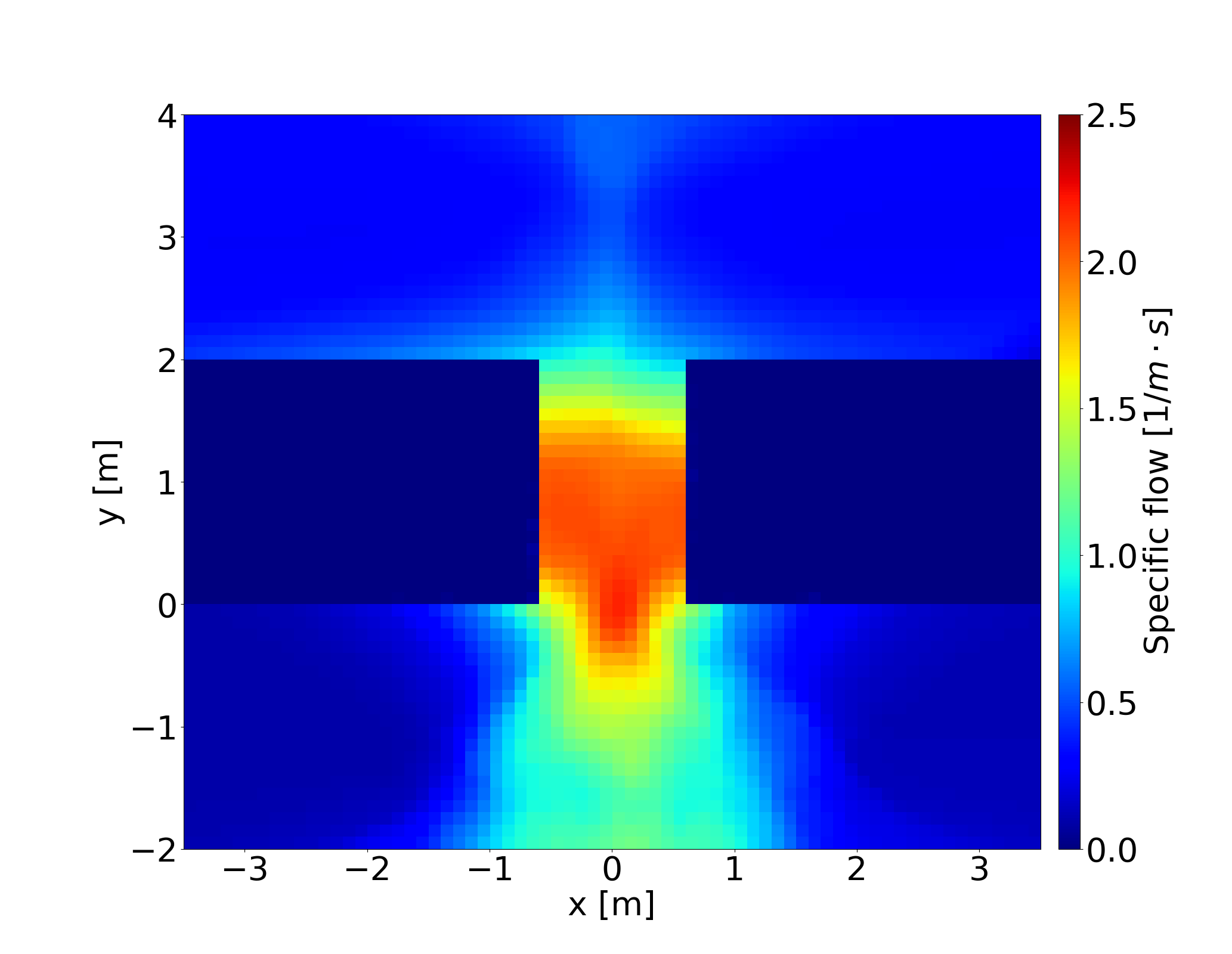}}
    \subfigure[Specific flow, original model]{\label{fig:subfig:h}
    \includegraphics[width=0.3\linewidth]{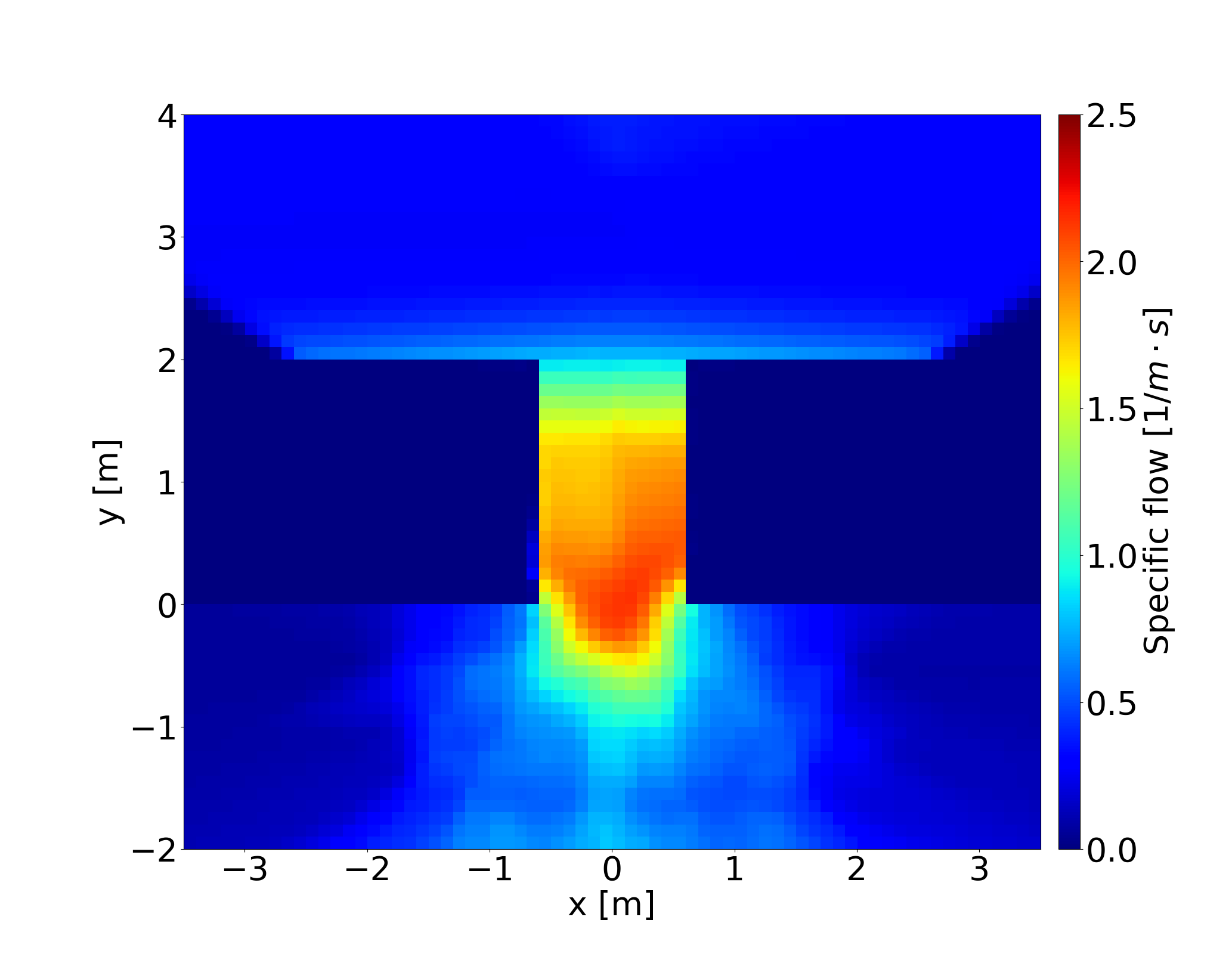}}
    \subfigure[Specific flow, new model]{\label{fig:subfig:i}
    \includegraphics[width=0.3\linewidth]{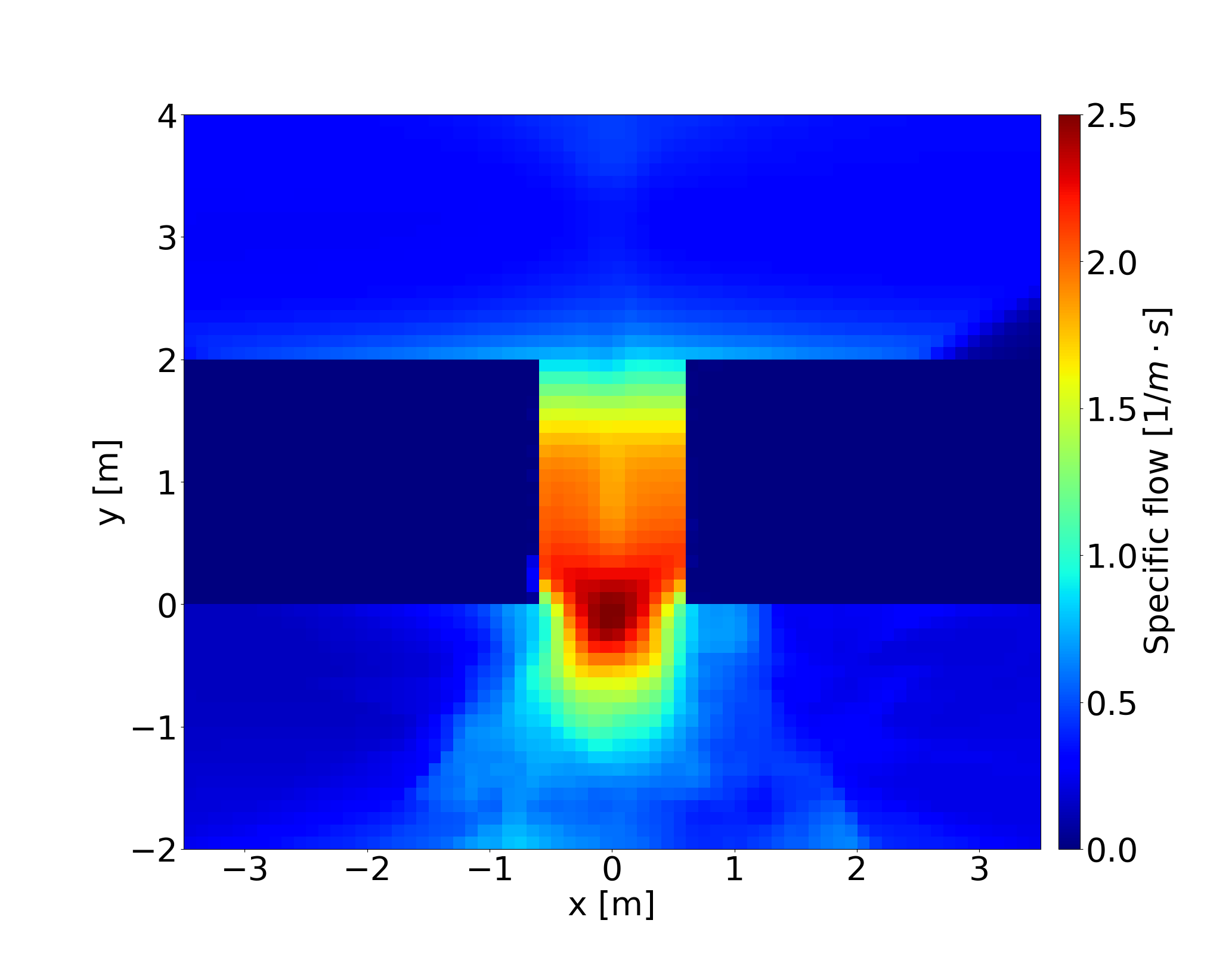}}
    \caption{Spatiotemporal profile of bottleneck flow, the width of bottleneck is 1.2~m and pedestrians through bottleneck from bottom to top.}
    \label{fig:profile}
\end{figure}

Finally, we compare the spatiotemporal profile of bottleneck flow when the width is 1.2~m. 
In simulations, we let pedestrians appear in same positions and at same times as in the experiment, in order to eliminate the impact of pedestrians' initial distribution. 
The profiles obtained from experiment, original speed model and new model are shown in figure \ref{fig:profile}.
Although profiles obtained from new model are still somewhat different from the experimental results, a visible enhancement can be observed. 
The pedestrians do not deviate strongly from the exit as it can be observed with the original model.

\section{Conclusion}
\label{sec:conclusion}
In this paper we generalize the collision-free model~\cite{vm2Cvm} by introducing several enhancements and new components that lead to better dynamics.
We firstly complete the collision-free velocity model by introducing the influence of walls.
Then, we generalize the definition of the model in order to consider dynamical ellipse shapes of pedestrian's projection on 2D space, instead of the circular one. 
Hereby, we define the semi-axes of the ellipses such that the two-dimensional fundamental diagram is well reproduced with respect to experimental data. 
After introducing a new direction sub-model, we show quantitatively that the unrealistic behavior of the agents during simulations with the original model could be mitigated.
Simulation results show that the new direction sub-model can remove unrealistic backward movement and undesired shaking behaviours without breaking the advantages of the original model.

Our validation of the model was systematic, going from the fundamental diagram in narrow corridors (1D) through fundamental diagrams in wide corridors (2d) to the flow-width relation in bottlenecks.
Although the generalized model produces better results, there are still some problems that have not been solved yet. 
First of all, jamming arch appears in bottleneck scenarios with a small width. 
Here, the collision-free nature of the model favours excessive blocking of agents in front of the exit.
Further investigations are necessary to investigate an appropriate mechanism for dealing with arching.
Besides, more detailed validations will be done in future work.  

\section*{Acknowledgments}
Qiancheng Xu thanks the funding support from the China Scholarship Council.

\bibliographystyle{elsarticle-num}

\bibliography{reference}

\begin{thebibliography}{10}
\expandafter\ifx\csname url\endcsname\relax
  \def\url#1{\texttt{#1}}\fi
\expandafter\ifx\csname urlprefix\endcsname\relax\def\urlprefix{URL }\fi
\expandafter\ifx\csname href\endcsname\relax
  \def\href#1#2{#2} \def\path#1{#1}\fi

\bibitem{review1}
A.~Seyfried, O.~Passon, B.~Steffen, M.~Boltes, T.~Rupprecht, W.~Klingsch, New
  insights into pedestrian flow through bottlenecks, Transportation Science
  43~(3) (2009) 395--406.

\bibitem{review2}
M.~Chraibi, A.~Tordeux, A.~Schadschneider, A.~Seyfried, Modelling of pedestrian
  and evacuation dynamics, Encyclopedia of Complexity and Systems Science
  (2018) 1--22.

\bibitem{review3}
D.~C. Duives, W.~Daamen, S.~P. Hoogendoorn, State-of-the-art crowd motion
  simulation models, Transportation research part C: emerging technologies 37
  (2013) 193--209.

\bibitem{macro1}
A.~Hanisch, J.~Tolujew, K.~Richter, T.~Schulze, Modeling people flow: online
  simulation of pedestrian flow in public buildings, in: Proceedings of the
  35th conference on Winter simulation: driving innovation, Winter Simulation
  Conference, 2003, pp. 1635--1641.

\bibitem{macro2}
R.~L. Hughes, A continuum theory for the flow of pedestrians, Transportation
  Research Part B: Methodological 36~(6) (2002) 507--535.

\bibitem{macro3}
R.~Hughes, The flow of large crowds of pedestrians, Mathematics and Computers
  in Simulation 53~(4-6) (2000) 367--370.

\bibitem{macro4}
R.~L. Hughes, The flow of human crowds, Annual review of fluid mechanics 35~(1)
  (2003) 169--182.

\bibitem{meso0}
L.~F. Henderson, On the fluid mechanics of human crowd motion, Transportation
  Research 8 (1974) 509--515.

\bibitem{meso1}
N.~Bellomo, A.~Bellouquid, On the modelling of vehicular traffic and crowds by
  kinetic theory of active particles, in: Mathematical modeling of collective
  behavior in socio-economic and life sciences, Springer, 2010, pp. 273--296.

\bibitem{meso2}
C.~Dogbe, On the modelling of crowd dynamics by generalized kinetic models,
  Journal of Mathematical Analysis and Applications 387~(2) (2012) 512--532.

\bibitem{meso3}
D.~Helbing, A fluid dynamic model for the movement of pedestrians, Complex
  Systems 6 (1998) 391--415.

\bibitem{ca1}
V.~J. Blue, J.~L. Adler, Cellular automata microsimulation for modeling
  bi-directional pedestrian walkways, Transportation Research Part B:
  Methodological 35~(3) (2001) 293--312.

\bibitem{ca2}
C.~Burstedde, K.~Klauck, A.~Schadschneider, J.~Zittartz, Simulation of
  pedestrian dynamics using a two-dimensional cellular automaton, Physica A:
  Statistical Mechanics and its Applications 295~(3-4) (2001) 507--525.

\bibitem{ca3}
M.~Fukui, Y.~Ishibashi, Self-organized phase transitions in cellular automaton
  models for pedestrians, Journal of the physical society of Japan 68~(8)
  (1999) 2861--2863.

\bibitem{ca4}
A.~Kirchner, A.~Schadschneider, Simulation of evacuation processes using a
  bionics-inspired cellular automaton model for pedestrian dynamics, Physica A:
  statistical mechanics and its applications 312~(1-2) (2002) 260--276.

\bibitem{ca5}
M.~Muramatsu, T.~Irie, T.~Nagatani, Jamming transition in pedestrian counter
  flow, Physica A: Statistical Mechanics and its Applications 267~(3-4) (1999)
  487--498.

\bibitem{vm1}
A.~Tordeux, A.~Seyfried, Collision-free nonuniform dynamics within continuous
  optimal velocity models, Physical Review E 90~(4) (2014) 042812.

\bibitem{vm2Cvm}
A.~Tordeux, M.~Chraibi, A.~Seyfried, Collision-free speed model for pedestrian
  dynamics, in: Traffic and Granular Flow'15, Springer, 2016, pp. 225--232.

\bibitem{vm3}
B.~Maury, J.~Venel, A discrete contact model for crowd motion, ESAIM:
  Mathematical Modelling and Numerical Analysis 45~(1) (2011) 145--168.

\bibitem{vm4}
S.~Paris, J.~Pettr{\'e}, S.~Donikian, Pedestrian reactive navigation for crowd
  simulation: a predictive approach, in: Computer Graphics Forum, Vol.~26,
  Wiley Online Library, 2007, pp. 665--674.

\bibitem{sfm1}
D.~Helbing, P.~Molnar, Social force model for pedestrian dynamics, Physical
  review E 51~(5) (1995) 4282.

\bibitem{mohcine}
M.~Chraibi, A.~Seyfried, A.~Schadschneider, Generalized centrifugal-force model
  for pedestrian dynamics, Physical Review E 82~(4) (2010) 046111.

\bibitem{sfm3}
D.~R. Parisi, M.~Gilman, H.~Moldovan, A modification of the social force model
  can reproduce experimental data of pedestrian flows in normal conditions,
  Physica A: Statistical Mechanics and its Applications 388~(17) (2009)
  3600--3608.

\bibitem{sfm4}
A.~Johansson, D.~Helbing, P.~K. Shukla, Specification of the social force
  pedestrian model by evolutionary adjustment to video tracking data, Advances
  in complex systems 10~(supp02) (2007) 271--288.

\bibitem{ellipse}
J.~J. Fruin, Pedestrian planning and design, Tech. rep., New York: Elevator
  World (1971).

\bibitem{aminT}
A.~Seyfried, B.~Steffen, W.~Klingsch, M.~Boltes, The fundamental diagram of
  pedestrian movement revisited, Journal of Statistical Mechanics: Theory and
  Experiment 2005~(10) (2005) P10002.

\bibitem{1DdataBottleneck}
A.~Seyfried, M.~Boltes, J.~K{\"a}hler, W.~Klingsch, A.~Portz, T.~Rupprecht,
  A.~Schadschneider, B.~Steffen, A.~Winkens, Enhanced empirical data for the
  fundamental diagram and the flow through bottlenecks, in: Pedestrian and
  Evacuation Dynamics 2008, Springer, 2010, pp. 145--156.

\bibitem{2Ddata}
S.~Holl, A.~Seyfried, Hermes-an evacuation assistant for mass events, Inside
  7~(1) (2009) 60--61.

\bibitem{Weidmann}
S.~Buchm{\"u}ller, U.~Weidmann, Parameters of pedestrians, pedestrian traffic
  and walking facilities, IVT Schriftenreihe 132 (2006).

\end{thebibliography}
\end{document}